\documentclass[12pt,letterpaper]{article}
\usepackage[utf8]{inputenc}
\usepackage{graphicx}
\usepackage{booktabs}
\usepackage{amsmath}
\usepackage{amsfonts}
\usepackage{amssymb}
\usepackage{array}     
\usepackage{enumitem}
\usepackage{caption}
\usepackage{subcaption}
\usepackage[toc,page]{appendix}
\usepackage[letterpaper,top=2cm,bottom=2cm,left=2cm,right=2cm]{geometry}
\usepackage[
backend=biber,
style=apa,
]{biblatex}
\usepackage{hyperref}
\hypersetup{
    colorlinks=true,
    linkcolor=blue,          
    citecolor=blue,          
    filecolor=blue,          
    urlcolor=blue            
}

\usepackage{subcaption}
\title{How Divorce Reforms Induced Married Couples to Supply More Labor\thanks{For helpful feedback, I thank my dissertation committee Ewout Verriest, Bradley Setzler, and Marc Henry. I also thank Andres Aradillas-Lopez, Vijay Krishna, Conor Ryan, Kala Krishna, the participants in the Penn State Econometric Reading Group for helpful comments.}}
\renewcommand{\thefootnote}{\fnsymbol{footnote}} 

\author{Yedilkhan (Eddie) Baigabulov \footnote{PhD Candidate, Department of Economics, Pennsylvania State University} }

\date{\today}

\begin{document}

\maketitle

\noindent 
\begin{center}
    \href{https://drive.google.com/file/d/1Yn8xYFjcfOw-bvQ729uoxoqVxN1LHXmu/view?usp=share_link}{[for the latest version, click here]}
\end{center}
\begin{abstract}
\textit{This paper studies the dynamic effects of divorce legislation on the labor supply behavior and welfare of married couples. Using U.S. household panel data, I examine the transition from the ``Title-based Regime” (TBR) in the 1970s to the ``Equitable Distribution Regime’’ (EDR) in the 1980s. 
Under the former regime, marital assets are split based on formal ownership rights, allowing spouses to commit ex-ante to a future asset allocation in the event of divorce. 
The shift to EDR, where courts more equitably divide marital assets, 
interacted in two ways with household uncertainty: First, because the court chooses the division of assets,  households  no longer can commit to a future asset allocation. Second, because the court's decision is  not fully predictable, the policy introduces a new source of uncertainty.
 For these reasons,  EDR increased the incentives for both spouses to self-insure by working to accumulate precautionary savings during marriage. Using a staggered difference-in-differences design, I confirm empirically that both married men and women increased labor supply in response to the policy change, with the women's increase equal to 30\% of the total rise in female labor force participation over the last 70 years. To rationalize and understand the welfare implications of the empirical evidence, I develop and estimate a dynamic model of household labor supply, savings and divorce.  I use the estimated model to show that the EDR may have inadvertently reduced  the welfare of both spouses even while married. }

\end{abstract}

\vfill

\renewcommand{\thefootnote}{\arabic{footnote}} 
\setcounter{footnote}{0} 
\clearpage

\section{Introduction}

This research delves into the social and economic outcomes stemming from the adoption of the Equitable Distribution Regime (EDR), a policy related to the redistribution of property upon divorce, and its subsequent influence on behaviors within households.

Traditionally, if a home bought during the marriage was only in the husband's name, it would stay his even after the divorce. The EDR was introduced as a corrective measure, endeavoring to create a fairer division of property, especially in favor of the second earner – often the wife – who, under the prior Title-Based Regime (TBR), frequently received a minimal share of assets upon divorce. The EDR mandates that spouses collaborate to determine the division of their shared property during divorce. In scenarios where mutual agreement is unattainable, the court intervenes, establishing a rule and deciding the property allocation. The court's decision is highly unpredictable, as it can be influenced by various factors, such as each spouse’s contributions to the household, income disparities, the number of children involved, and others. Although the precise decision-making criteria might not be explicit, existing literature, like \textcite{woodhouse2006divorce}, proposes that the primary earner, often the husband, might receive between half and two-thirds of the assets. Moreover, \textcite{garrison1991good}, who analyzed the court outcomes before and after the adoption of EDR in New York, showed that the court decision indeed led to a higher share for the wife than before the policy adoption (see Appendix \ref{sec: Asset distibution NY}).

\begin{figure}[htbp!]
    \centering
    \caption{Female Labor Participation Changes: 1950-2019 Change vs EDR Effect}
    \includegraphics[width=0.6\textwidth]{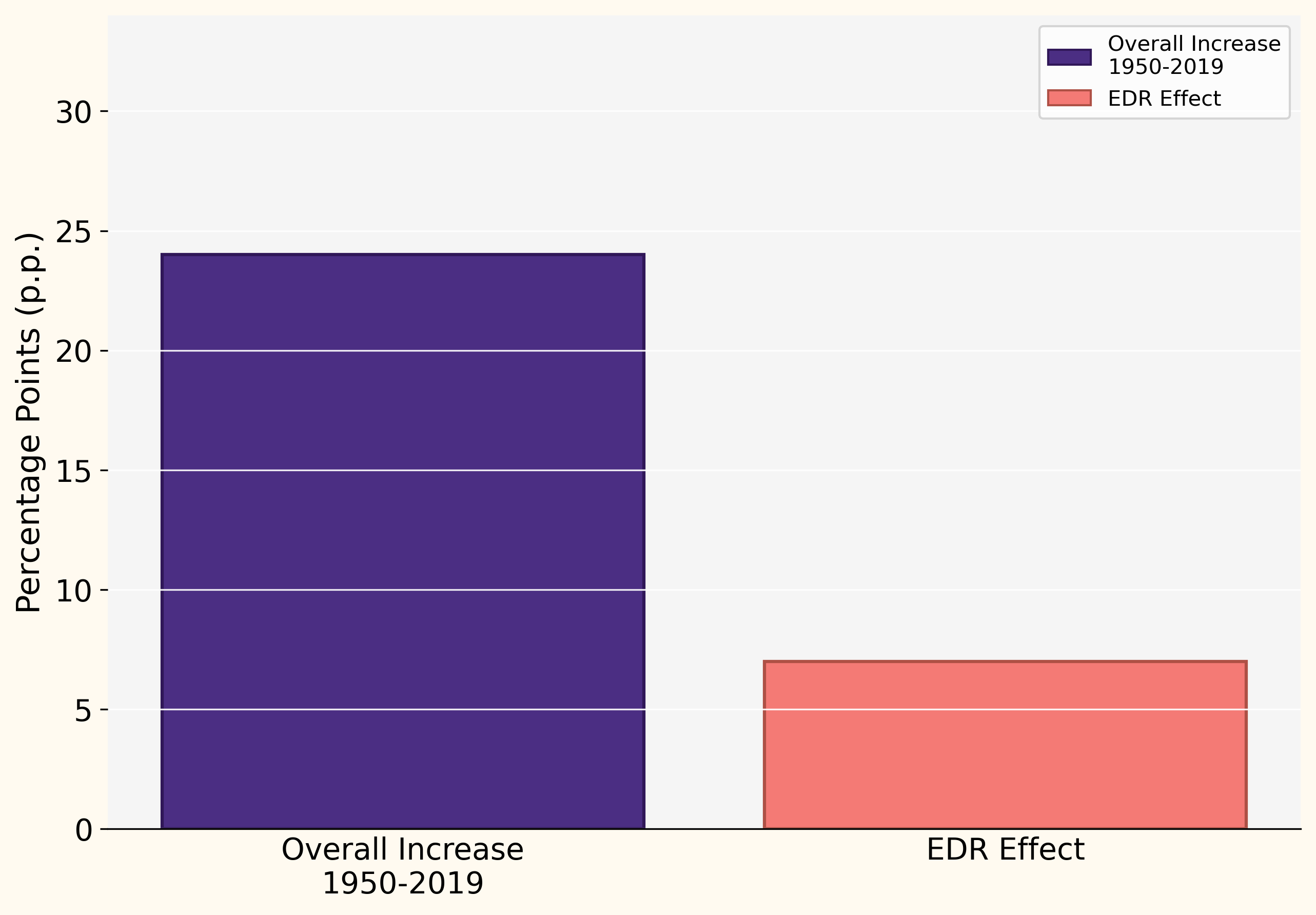}
    \label{fig:female_labor_20th}
    \caption*{\footnotesize \textit{Note}: The overall increase is computed based on historical data from the Federal Reserve Bank of St. Louis.}
\end{figure}

To examine EDR's impact, this study leverages policy variations across states, using the recent difference-in-differences (DiD) methods (\textcite{callaway2021difference}, \textcite{borusyak2021revisiting}) to address staggered adoption bias. Findings indicate that EDR led to higher labor supply of both spouses and savings.  Similar to \textcite{manski2018right}, I show that the results are robust to significant violation of parallel trend assumption.The significance of this finding could be underscored by the fact that the effect of EDR on labor supply of wives is equal to 30\% of overall increase female labor force participation from 1950 to 2019 (See Figure \ref{fig:female_labor_20th}).

Even if a divorce is a collective decision, it is driven by unexpected shocks (\textquotedblleft love\textquotedblright shock) and is treated as such by the household. The following mechanism explains how EDR might have propelled higher labor supply among women: Under the TBR, households could commit to post-divorce property shares by formally titling the respective assets. This commitment device allowed for the elimination of the economic consequences of divorce on consumption by ensuring complete smoothing across different marital statuses in the future e.g. if a spouse's future consumption is 2 units when staying married, it will remain 2 units in the case of a divorce. But, EDR eliminated this ability of the household to choose the share of property of each spouses upon divorce, replacing it with court-decided allocations, subsequently transforming divorce from fully-insured to uninsurable risk. Households, typically risk-averse, respond to higher uncertainty (risk) with higher savings via either reduced current consumption or increased labor supply (or both).
The EDR policy could drive precautionary savings in two main ways:
\begin{itemize}
\item[a)]  \textit{Deviation from optimal share}:  Even with a fixed, commonly known property split, precautionary savings can be triggered if the enforced rule is misaligned with optimal choices made under the TBR. For example, under TBR, households can offset consumption fluctuations post-divorce by choosing asset divisions that ensure smooth consumption for both spouses regardless of marital status. However, if a new policy mandates equal asset division—analogous  to the Community Property regime in states like California and Texas—and this enforced split is different from their original optimal choice under TBR, it fails to eliminate consumption shocks following a divorce (consumption is different across marital statuses). So "prudent" households will increase savings in response to EDR, which could be achieved by higher labor supply.
\item[b)] \textit{Random share}: While even a deterministic property division (from (a)) could lead to higher labor supply from both spouses, the actual EDR introduces an additional element of uncertainty due to potential court intervention in asset allocation. This uncertainty, conjoined with the initial risk of divorce, further strengthens precautionary savings incentives.
\end{itemize}

In a two-period household model, I derive the conditions under which the EDR increases savings, drawing significantly upon the concept of \textit{relative prudence}. Prudence plays a pivotal role in the precautionary saving literature, as it defines an agent's incentives to increase or decrease savings in response to higher uncertainty. Drawing parallels with the findings by \textcite{vergara2017precautionary}, I establish a condition that, while sufficient for general utility functions, holds as both necessary and sufficient for special case of the CRRA utility function, \(\frac{c^{1-\gamma}}{1-\gamma}\) . This leads to the finding that, in a CRRA-based model, the \textbf{only} parameter determining the sign of EDR's impact on labor and savings is relative prudence. This parameter is  constant for CRRA utility case and equal to \(1 + \gamma\) and the treatment effect of EDR on labor supply changes sign when relative prudence, $1+\gamma$, equals 2. This suggests that precautionary saving is the only plausible explanation for the positive effect of EDR on labor supply. This remains the sole mechanism in numerous extensions, as the sign of the treatment effect consistently switches at relative prudence equal to 2, even when I introduce  an endogenous divorce and the ability of spouses to ``bribe" (bargaining) one another in cases where only one spouse desires a divorce. Transitioning from theoretical analysis to empirical implementation, I  estimate  a structural household model which is estimated by matching number of data moments including the ATT produced by staggered DiD. Welfare analysis reveals that despite the policy's aim to enhance the welfare of wives, the actual impact of the policy proves detrimental to the welfare of both spouses. This counter-intuitive outcome can be attributed to the fact that while wives may secure a larger share in the event of a divorce, they, alongside their husbands, experience decreased welfare during the marriage. This downturn arises due to the elevated uncertainty, viewed from the household's perspective, which leads to a reduction in leisure and consumption during the marriage. Consequently, the diminished utility experienced during the marriage outweighs any potential utility increase that might occur in the event of a divorce. An important implication is that a policy which simply splits property equally - analogous to Community property regime in states like California, Texas etc - can provide wives the same share they would get on average under EDR, but would not result in such a significant welfare loss.

 This research provides a new perspective on the factors contributing to the rise in female employment, challenging the traditional narrative that primarily attributes this increase to decreasing gender inequality and evolving social roles for women. It suggests that a substantial part of this increase can be explained by the adoption of EDR. These laws heightened the financial uncertainty that families faced, often leaving them with no option but to increase their labor supply. This observation is not confined solely to the U.S.; several other countries either transitioned from TBR to Community property regimes or to EDR (such as Canada, the UK, Czech Republic, etc.) or moved from community property regimes to EDR \footnote{Most countries maintained the Community property regime as an optional choice for couples upon marriage. However, EDR became the default selection.} (examples include Sweden, Denmark, Norway, South Africa, and others) during the 20th century\footnote{It’s important to understand that property laws differ widely, even among states in the U.S. In the context of various countries, I refer to EDR as the policy that does not necessarily divide property evenly but permits some degree of court intervention.}.

Previous research in the domain of social and economic impacts of changes in divorce laws, such as work by \textcite{stevenson2007marriage} and \textcite{gray1998divorce}, has delved into aspects of marital investments and the labor decisions of married women in light of new divorce laws, uncovering  diminished investment in marital assets and shifts in labor-supply behaviors.

 Recent work by \textcite{voena2015yours} and \textcite{fernandez2017free} analysed the effect of Unilateral Divorce (UD) on household behavior depending on underlying property regime. \textcite{voena2015yours} demonstrated that the adoption of UD, particularly in states with equal property division, led to increased savings and a decline in female labor supply. Conversely, \textcite{fernandez2017free} explored the diverse impacts of  UD, uncovering varied benefits among different socioeconomic groups. This paper shifts focus towards the EDR, abstracting  from the widely-studied UD states and introducing a modified empirical approach for causal effect estimation. While \textcite{voena2015yours} employed regular OLS methods, which produce biased results in a staggered adoption setting, this study employs a DiD method robust to staggered adoption, as proposed by \textcite{callaway2021difference}. This approach illuminates a significant, positive effect of EDR on female labor supply and savings, contrasting with the findings of \textcite{voena2015yours}, who reported non-significant and negative effects on these respective variables.

This paper posits a central hypothesis that under the EDR, divorce is associated with  higher uncertainty because households cannot mitigate the impact on consumption by endogenously selecting the property share for each spouse — what is achievable under the TBR by assigning formal ownership for each property unit. Consequently, due to this increased uncertainty under EDR, households exhibit an inclination to save more than under TBR. This logic in line with the existing literature on precautionary savings (\textcite{kimball1989precautionary};  \textcite{vergara2017precautionary} etc.). A notable segment of the existing literature, including  \textcite{carroll1998important} and \textcite{lusardi1997precautionary}, highlights the idea that households commonly use savings as a financial safeguard. Furthermore, \textcite{blundell2008consumption} and \textcite{kaplan2010much} study how households utilize precautionary savings to insure themselves against both permanent and temporary shocks. \textcite{blundell2008consumption} conclude that precautionary savings can almost entirely insure against transitory shocks and also can be effective against permanent ones. This supports the hypothesis, suggesting that the financial uncertainty associated with EDR might trigger precautionary savings behavior and, consequently, influence household labor supply.

My paper also contributes to broader literature about post-divorce financial transfer and initial property division is only one example of such transfers. Another important type of such transfers are child support and alimony and there is large literature studying the implication of such support on spouses (\textcite{graham1989effect}; \textcite{ong2020effect}; \textcite{foerster2021untying}; \textcite{friday2022monetary} etc). One recent example, \textcite{foerster2021untying} shows that alimony is associates with higher labor supply disincentives compared to child support payments and is less efficient in providing consumption insurance.

The paper is organized as follows: Section \ref{sec: institute and data} explains the institutional background and introduces the data used in the research. Section \ref{sec: Emp strat and res} outlines the empirical strategy and presents the reduced-form results. Section \ref{sec: model} introduces the theoretical model and provides the conditions under which EDR stimulates precautionary savings motives. Section \ref{sec: Econometric and identification} describes the econometric implementation used for model estimation, presents parameter estimates, and Section \ref{sec: Counteractual} utilizes the estimated parameters for counterfactual simulations. Section \ref{sec: conclusion} concludes

\section{Institutional Context and Data Sources.} \label{sec: institute and data}
\subsection{Divorce Reform in the United States}
\begin{figure}[h!]
    \centering
    \caption{Transition dynamics of states between different divorce laws}\label{fig: property states}
    \includegraphics[scale=0.25]{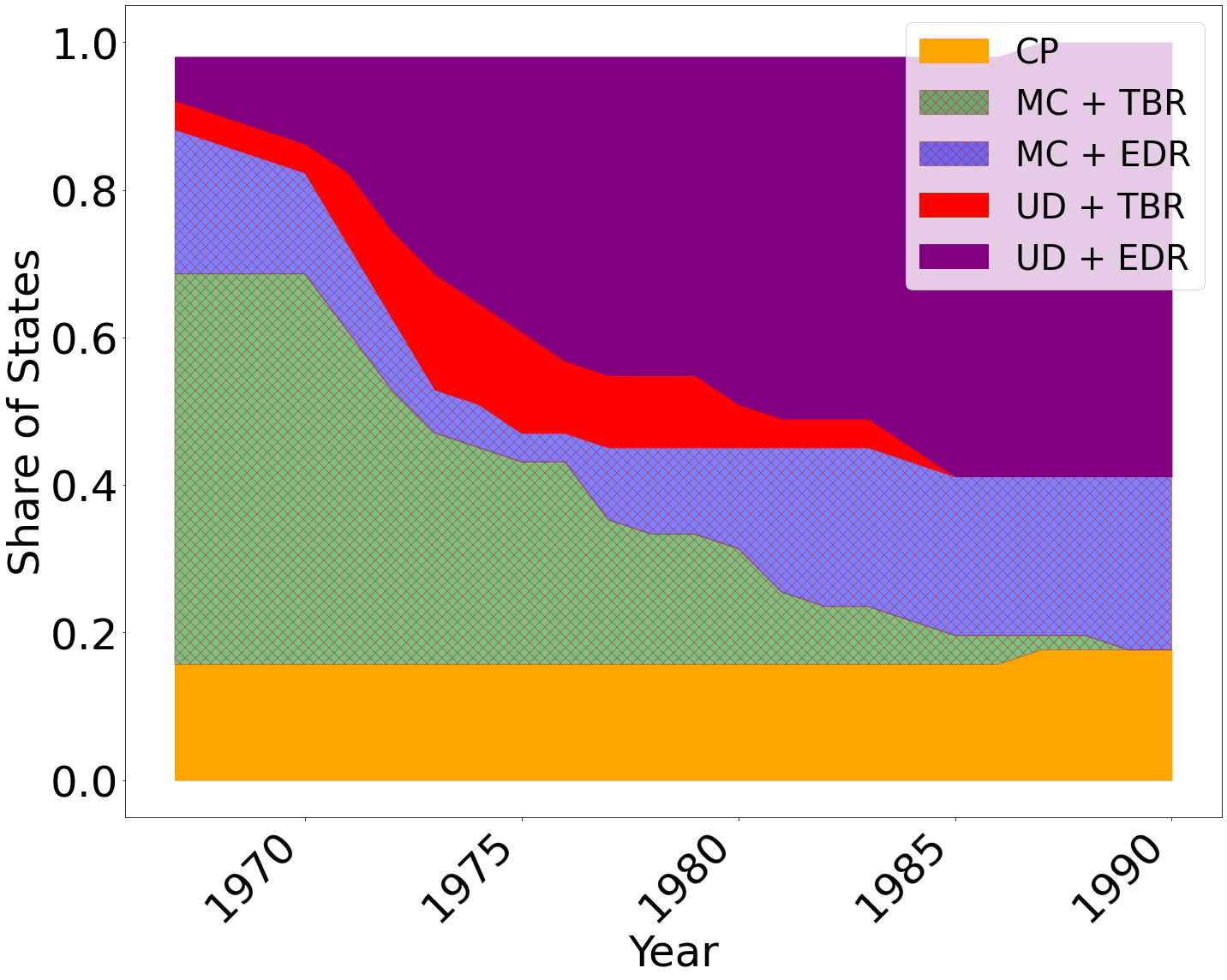}
    \caption*{\footnotesize \textit{Note}: States transitioned either from the Title-based property regime to the Community Property regime or to the Equitable Distribution regime. There were no transitions between the Equitable Distribution and Community Property regimes. Abbreviations: CP - Community Property regime, MC - Mutual Consent, TBR - Title-Based property regime, EDR - Equitable Distribution regime, UD - Unilateral Divorce.}
\end{figure}

From the 1960s to the late 1980s, numerous states transitioned from a Title-based property regime (TBR) to an Equitable Distribution Regime (EDR). In the former system, upon divorce, each spouse retained the property officially registered under their name. This often favored men, as they traditionally held ownership over the majority of assets. Conversely, EDR allows spouses to determine the division of property acquired during the marriage upon divorce. In cases of disagreement, the court intervenes, aiming to ensure a more equitable distribution of assets upon the dissolution of marriage. As evidence of the policy's effectiveness in increasing the share of property received by wives, I refer to actual court outcomes in New York before and after the adoption of EDR (see Table \ref{tab: asset distribution law } and Table \ref{tab: asset_type_distribution_law}). It can be observed that courts indeed started allocating a higher share of property to wives. Another regime, the Community Property (CP), mandates an equal distribution of property acquired during marital tenure. Most states that adopted this approach\footnote{For instance, Wisconsin adopted the Community Property regime in 1987.} did so before the datasets available for my analysis were created. Thus, my study is limited to states that adopted the Equitable Distribution regime. Figure \ref{fig: property states} demonstrates how the share of states under different regimes changed over time.

A significant legal alteration during this period was the introduction of Unilateral Divorce (UD), which replaced the Mutual Consent (MC) regime.
Figure \ref{fig: property states} plots the share of states with UD as the red and purple areas. My analysis is limited to states that did not transition to UD between 1968 and 1990, which is the \textit{subset} of the green and blue areas\footnote{Green and blue areas include states that did adopt UD at some point between 1968 and 1990. For example, Colorado until 1971 was under MC and TBR (green area in Figure \ref{fig: property states}) but in 1971, it adopted both UD and EDR, moving it to the purple area. Since it adopted UD within this time window, it is excluded from my analysis.}. These states include South Carolina, Maryland, Missouri, District of Columbia, Virginia, North Carolina, Arkansas, Mississippi, New York, Tennessee, New Jersey, Illinois. Limiting the scope to these states prevents the potential confounding influence of the two policies. I do not demonstrate how CP regime states transitioned to UD because, firstly, those states are not considered in my research, and secondly, it would render Figure \ref{fig: property states} difficult to interpret. 

As shown in Figure \ref{fig: property states}, all states eventually adopted EDR. Therefore, for my control group, I include states that had not yet adopted the policy in a given year. For example, Arkansas, which had not adopted EDR before 1977, can be included in the control group for states that adopted EDR prior to 1977. I also exclude Pennsylvania as it switched from the TBR Community Property regime and then reverted to TBR again during the first half of the 20th century. To my knowledge, it is the only example of this kind, making it an outlier. Wisconsin is also excluded from the control group because it is the only state that adopted CP during the 20th century, whereas the rest of the CP states adopted it much earlier, thus making it another outlier.

\subsection{Data Sources.} \label{sec:data}

This study utilizes data from the Panel Study of Income Dynamics (PSID) and the National Longitudinal Survey of Mature and Young Women (NLSW) datasets. The focus is exclusively on families that entered into marriage prior to the adoption of EDR, with the sample consisting of families in which both the wife and husband are aged between 22 and 65 years. States transitioning to the community property regime are excluded due to their transitions occurring before the first wave of the PSID. An exception is made for Wisconsin, which transitioned to the Community Property Regime in 1987. However, given that Wisconsin is the sole exception, it does not provide enough variation for causal analysis. Furthermore, states that adopted UD are omitted from the study. Notably, 15 states refrained from transitioning from mutual consent  to UD during the study period. Pennsylvania is also excluded due to its temporary transition to the Common Property Regime during the first half of the 20th century, potentially categorizing it as an outlier. Consequently, the focus is placed on the following states only: South Carolina, Maryland, Missouri, District of Columbia, Virginia, North Carolina, Arkansas, Mississippi, New York, Tennessee, New Jersey, and Illinois.

Table \ref{table: summary table} presents the descriptive statistics of the dataset. A key observation is that the employment rate for men under the TBR is nearly 90\%, indicating that the majority of them are employed. Regarding educational attainment, both men and women, on average, have completed high school.

Although the PSID provides the majority of the valuable information for the research, the NLSW is important as it contains asset information, which is a key variable for my analysis. A challenge with the NLSW is its irregular data collection, as the waves have significant gaps, limiting its utility for proper empirical analysis. For this reason, NLSW data is utilized for moments matching only. The variable 
$assets_t-assets_{t-1}$
  represents interpolated values due to the presence of gaps in the survey. It is calculated using the following formula:
\begin{equation}
\overline{assets_t-assets_{t-1}}=\frac{assets_t-assets_{t-r}}{r}
\end{equation}
where $t-r$ is the last time the woman was surveyed. For instance, I observe woman with total household savings \$2,000 in 1975 and observe her next time in 1978 with total household savings  \$ 3,000. Given that I have time gaps in survey I can not compute the annual change in savings dring this period so use the data for 1975 and 1978 to compute the average change in savings during this period $\frac{\$ 3000 - \$ 2000}{1978-1975} \approx \$ 333.3$. So I impute this value for the change between year 1978 and 1977 . 

 \begin{table}[h!]
        
         \caption{Summary Statistics}\label{table: summary table}
        \centering
        \begin{tabular}{lcccccc}
            \toprule
            & \multicolumn{3}{c}{Title Based} & \multicolumn{3}{c}{Equitable} \\
            & \multicolumn{3}{c}{Property Regime} & \multicolumn{3}{c}
            {Distribution Regime} \\
             \cmidrule(lr){2-4} \cmidrule(lr){5-7}
            &  \multicolumn{6}{c}{PSID} \\
           
            Variable & Mean & Std & N & Mean & Std & N \\
            \midrule
            Age of wife & 38.41 & 12.02 & 12,137 & 37.74 & 11.97 & 5,235 \\
              Age of husband & 43.15 & 13.26 & 12,137 & 42.30 & 13.19 & 5,235 \\
            Duration of marriage & 18.20 & 12.78 & 9,305 & 17.71 & 13.05 & 4,785 \\
            Wife Y. of Edu $>$ 12 & 0.24 & 0.42 & 12,137. & 0.39 & 0.49 & 5,235 \\
            Husband Y. of Edu $>$ 12 & 0.23 & 0.42 & 11,727 & 0.39 & 0.49 & 4,824 \\
            Work Hours (wife) & 15.96 & 16.67 & 12,137 & 20.54 & 17.13 & 5,235 \\
            Work Hours (husband) & 35.93 & 17.17 & 12,137 & 35.71 & 17.29 & 5,235 \\
            Employment (wife) & 0.60 & 0.49 & 12,137 & 0.69 & 0.46 & 5,235 \\
            Employment (husband) & 0.90 & 0.31 & 12,137 & 0.88 & 0.32 & 5,235 \\
            &  \multicolumn{6}{c}{NLSW} \\
           
            Variable & Mean & Std & N & Mean & Std & N \\
            \midrule
            $assets$ & 18,975.5 & 28,502.64 & 1,036 & 23,962.81 & 33,559.07 & 265 \\
              $ \overline{assets_t-assets_{t-1}}$ & 47.10 & 260.52 & 1,036 & 46.64 & 212.41 & 265 \\
              age & 33.87 & 9.15 & 1,036 & 50.02 & 5.64 & 265 \\
            \bottomrule
        \end{tabular}
    \end{table}

\section{Empirical Strategy and Results}\label{sec: Emp strat and res}

\subsection{DiD for Staggered Adoption}
The primary goal of this study is to investigate the impact of EDR adoption on the labor supply and savings of household. The variation in property law across states and over time provides an excellent context  for this investigation. The staggered adoption of the policy by different states\footnote{States implemented the policy at different times.} poses challenges to the straightforward OLS-based DiD approach. With the recent development in literature on staggered adoption, as noted in works like \textcite{callaway2021difference}, \textcite{borusyak2021revisiting}, \textcite{goodman2021difference}, \textcite{de2020two}, \textcite{sun2021estimating}, and others, it becomes evident that standard regressions might lead to biased outcomes. Consequently, I use the methodologies proposed by \textcite{callaway2021difference} (CS) and \textcite{borusyak2021revisiting} (BJS). Although both are robust to the bias arising from staggered adoption in the OLS setting, I  consider BJS as my main method due to its efficiency in producing estimates\footnote{As demonstrated later, both methods yield  similar results.}. The CS methodology is elaborated further in Appendix \ref{sec: Diff-in-Diff description}.

\subsubsection*{\textbf{Parallel Trend Assumption}}

Among the various assumptions underpinning the DiD approach, the \textit{parallel trend} assumption stands out as particularly critical and is hence addressed herein.

\begin{itemize}[label=$\diamond$]
    \item \textbf{Unconditional Parallel Trend (UPT).} There exist non-stochastic  $\alpha_i$ and $\beta_t$ such that
    \[
    \mathbb{E} \left[ Y_{it}(0) \right] = \alpha_i + \beta_t .
    \]
    This is the strongest assumption, prevalent in literature and  considered as the default in BJS. Its strength stems from assuming a parallel trend for each unit, irrespective of covariates. It implies, for example, that employment trends are parallel for both genders. If there is a clear parallel trend conditional on gender but not across genders, and if there's a significant difference in gender composition between the treated and control groups, then the UPT assumption fails, while the \textit{conditional} one holds.
    
    Note that, as also acknowledged by the authors, BJS yields unbiased results even if only the \textit{conditional} parallel trend assumption is satisfied.
    \item \textbf{Conditional Parallel Trend (CPT).}
    \begin{align}
    \mathbb{E}\left[Y_t(0) - Y_{t-1}(0) | X, G_g = 1\right] &= \mathbb{E}\left[Y_t(0) - Y_{t-1}(0) | X, D_s = 0, G_g = 0\right],
    \end{align}
    where $Y_t(0)$ is the potential untreated outcome, and $G_g$ and $D_s$ are indicators for treatment in periods $g$ and $s$, respectively. The covariates, represented by $X$, are determined before period $g$. This weaker assumption, compared to the UPT, allows for variation in trends conditional on covariates. Yet, given a set of covariates $X$, the trend has to be parallel. Authors like \textcite{manski2018right} and \textcite{rambachan2023more} in more recent research have relaxed this assumption to allow bounded variation in the trend.

    \item \textbf{Bounded Variation Assumption.} Using notation from CS, this assumption can be written as
    \[
    \left| \mathbb{E}\left[Y_t(0) - Y_{t-1}(0) | G_g = 1\right] - \mathbb{E}\left[Y_t(0) - Y_{t-1}(0) | D_s = 0, G_g = 0\right] \right| \leq \Delta,
    \]
    suggesting that while the trend might not be strictly parallel, the difference in trends should be bounded. The value of $\Delta$ is determined by the researcher. For this study, the largest observed deviation from zero before EDR adoption serves as the baseline for $\Delta$.
\end{itemize}
\subsection{Estimation Algorithm and Empirical Strategy}

Before delving into empirical findings, I provide an overview of the employed estimation algorithm and empirical strategy based on BJS, however, description of CS could be found in  Appendix \ref{sec: Diff-in-Diff description}.

\begin{enumerate}
    \item Combine all untreated ("not-yet treated") units and estimate:
    \begin{equation}
        Y = A' \lambda + X' \delta + \varepsilon.
    \end{equation}
    where $X$ includes regular covariates and $A$ consists of fixed effects
    \item For every treated unit, compute $\hat{\tau}_{i,t}=Y_{it} -\hat{Y}_{i,t}(0)$, where $\hat{Y}_{i,t}(0)=A' \hat{\lambda} + X' \hat{\delta}$.
    \item Apply $\hat{\tau}_{i,g+r}$ for different aggregations, as illustrated by BJS, to get an aggregate average treatment effect (ATE) for various periods post-intervention.
\end{enumerate}

\textbf{Fixed effects ($A$)}. Given that all my covariates are discrete I use fixed effects only without using regular controls. Along with year and individual fixed effects, I include fixed effects for age and education of both spouses, number of children, duration of marriage.

\subsection{Empirical Results.}

\subsubsection{Event Study and Parallel Assumption Test.}
\begin{figure}[h!]
\centering
\caption{Effects of EDR on Wife's Weekly Working Hours and Employment}\label{fig:wife_labour_supply_effect}
\begin{subfigure}{.5\textwidth}
\centering
\includegraphics[width=1\linewidth]{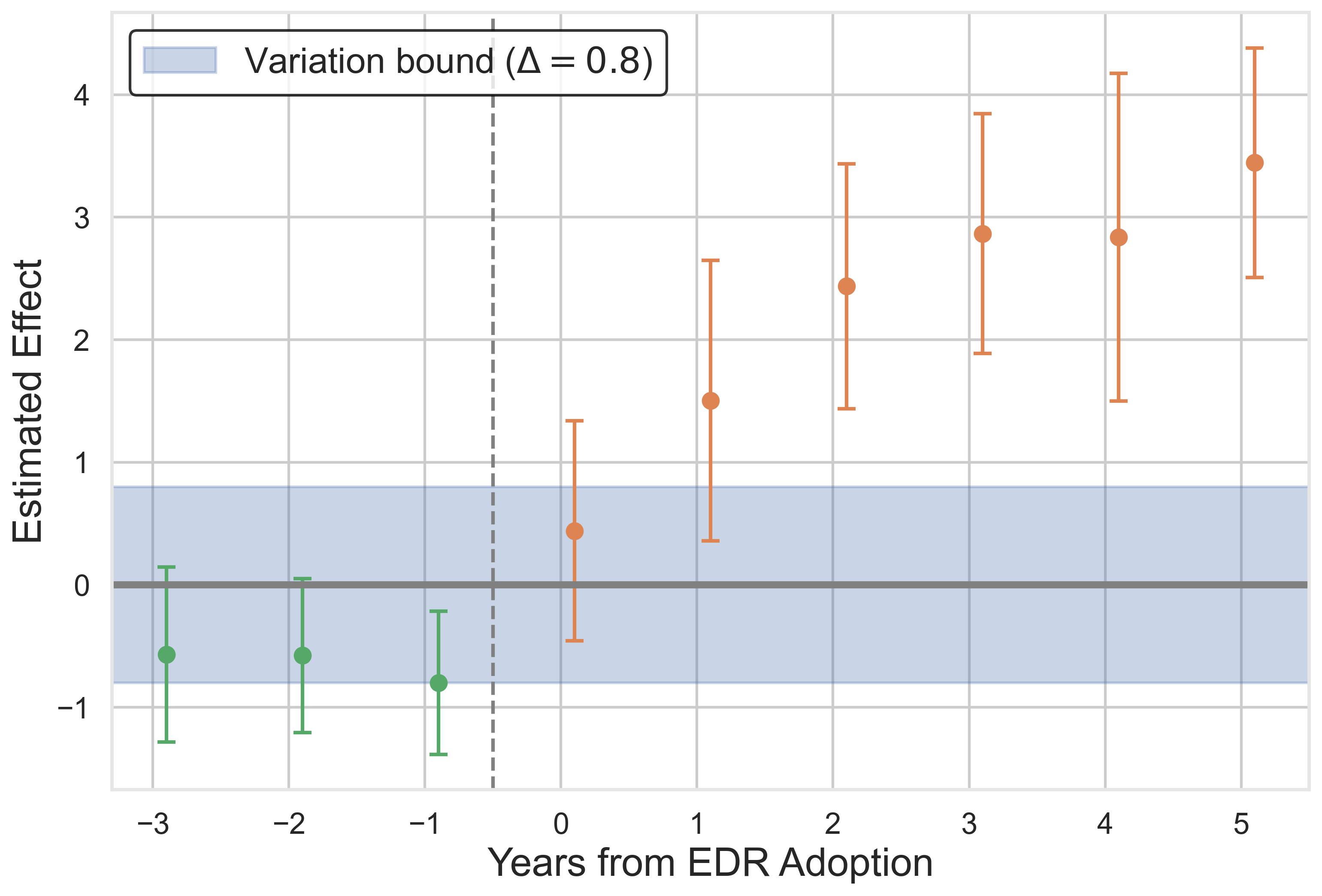}
\caption{) Weekly work hours}
\label{fig:wf_work_hours}
\end{subfigure}%
\begin{subfigure}{.5\textwidth}
\centering
\includegraphics[width=1\linewidth]{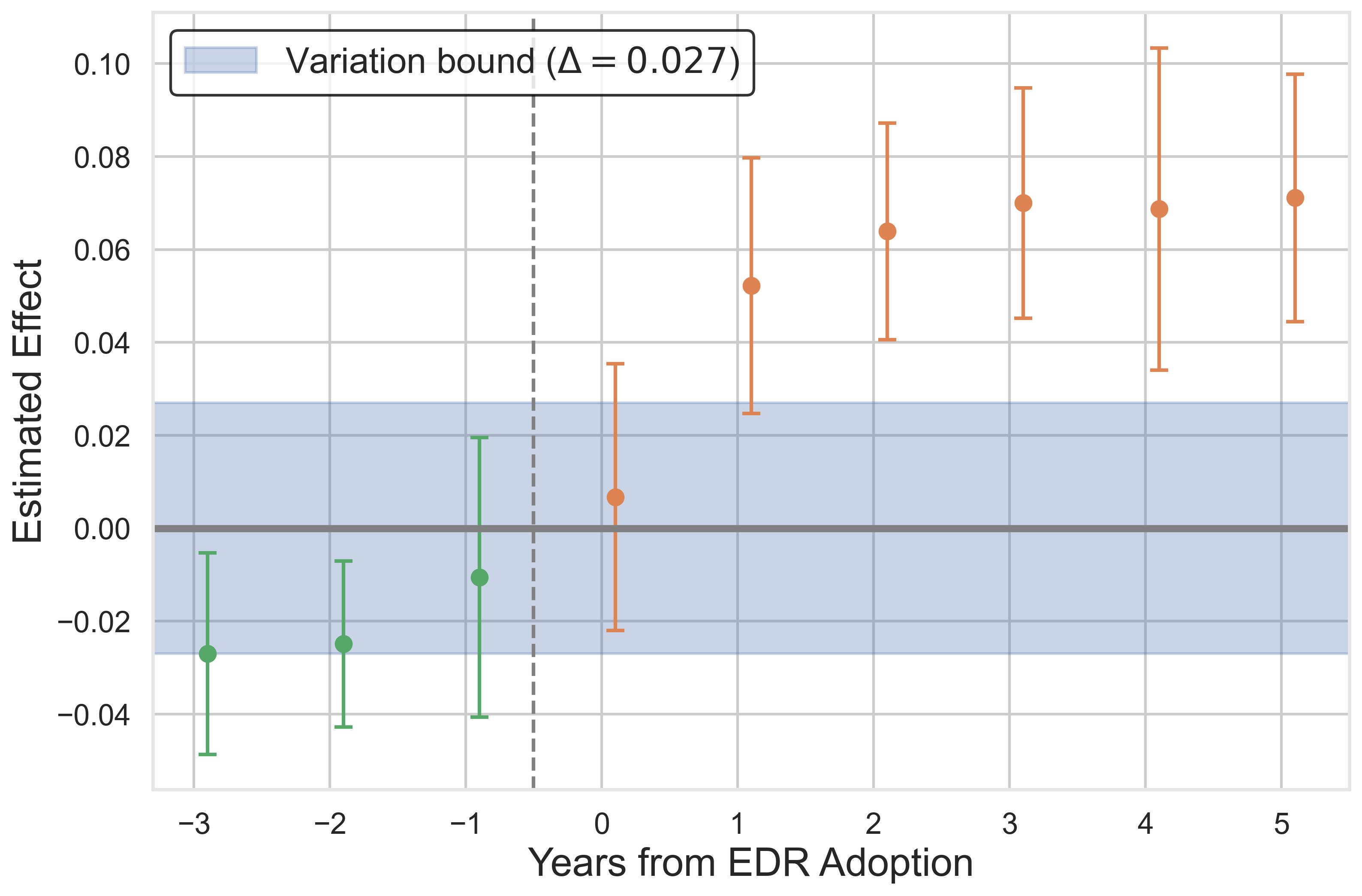}
\caption{) Employment (binary)}
\label{fig:wf_work_dummy}
\end{subfigure}
\caption*{\footnotesize \textit{Note}: \footnotesize   a) FE :   marriage duration, husband's age, wife's age,  N. of children, education of both spouses.  b) Confidence intervals have a 95\% coverage. c) Families observed for fewer than 5 years post-treatment are excluded from the treatment group to avoid attrition bias. They can, however, serve as control groups. d) Variation Bound uses the higher in absolute value coefficient from pre-treatment period as $\Delta$}

\end{figure}

From Figure~\ref{fig:wife_labour_supply_effect}, it is evident that adopting EDR influences the labor supply of married women. The effect stabilizes at an increase of approximately 3 hours in working hours and a 7 p.p. rise in employment. While Figure~\ref{fig:wife_labour_supply_effect} presents results produced by BJS, CS produces similar outcomes (see Figure~\ref{fig: both methods wife}). The significance of these findings is underscored by the fact that the female labor participation rate from 1950 to 2019 increased by 24 percentage points, meaning that the effect of EDR constitutes 30\% of the overall increase in female labor participation over the last 70 years.

\begin{figure}[h!]
\centering
\caption{Effects of EDR on Husband's Weekly Working Hours and Employment (dummy)}\label{fig:husband_labor_supply_effect}
\begin{subfigure}{.5\textwidth}
\centering
\includegraphics[width=1\linewidth]{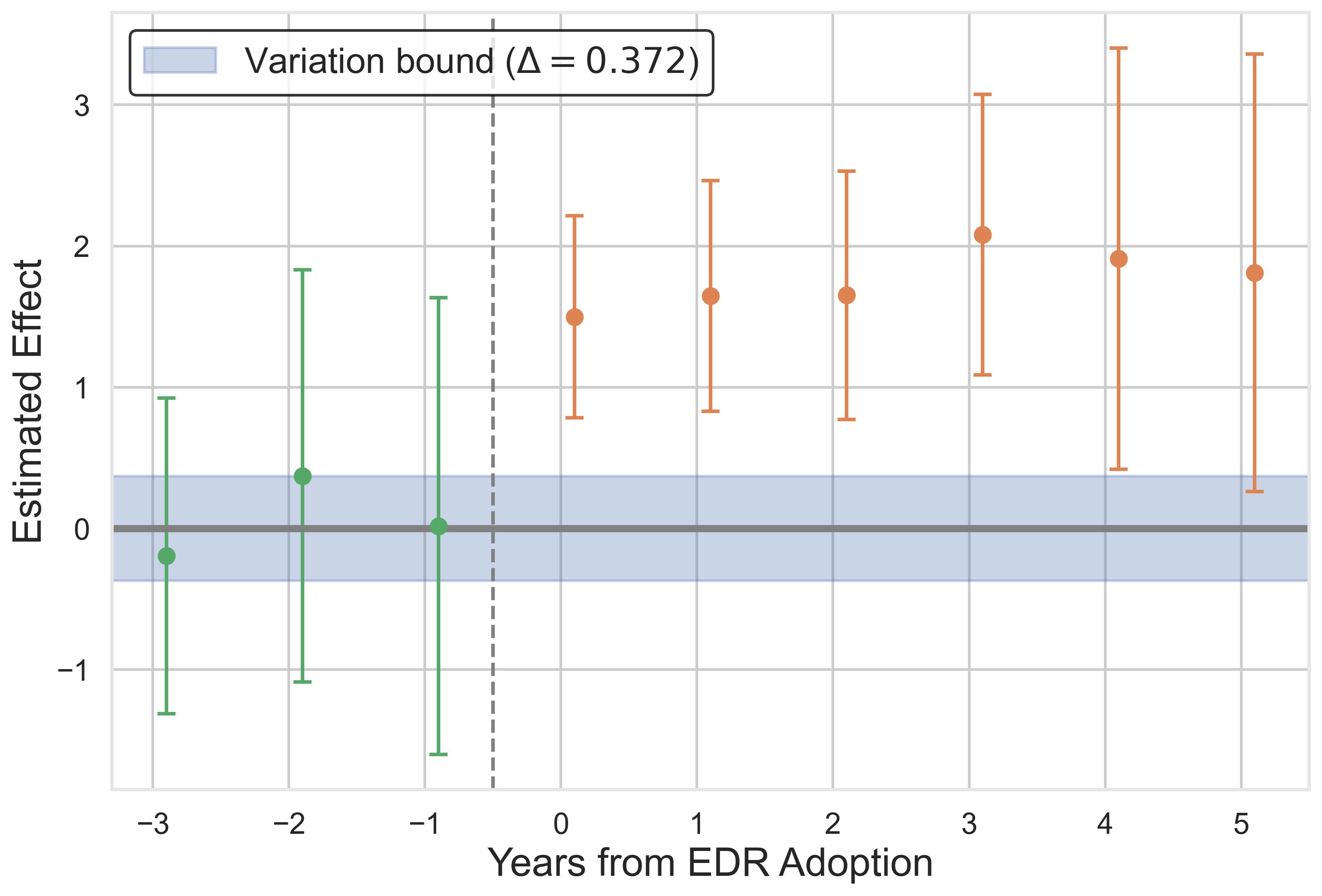}
\caption{) Weekly work hours}
\label{fig:hd_work_hours}
\end{subfigure}%
\begin{subfigure}{.5\textwidth}
\centering
\includegraphics[width=1\linewidth]{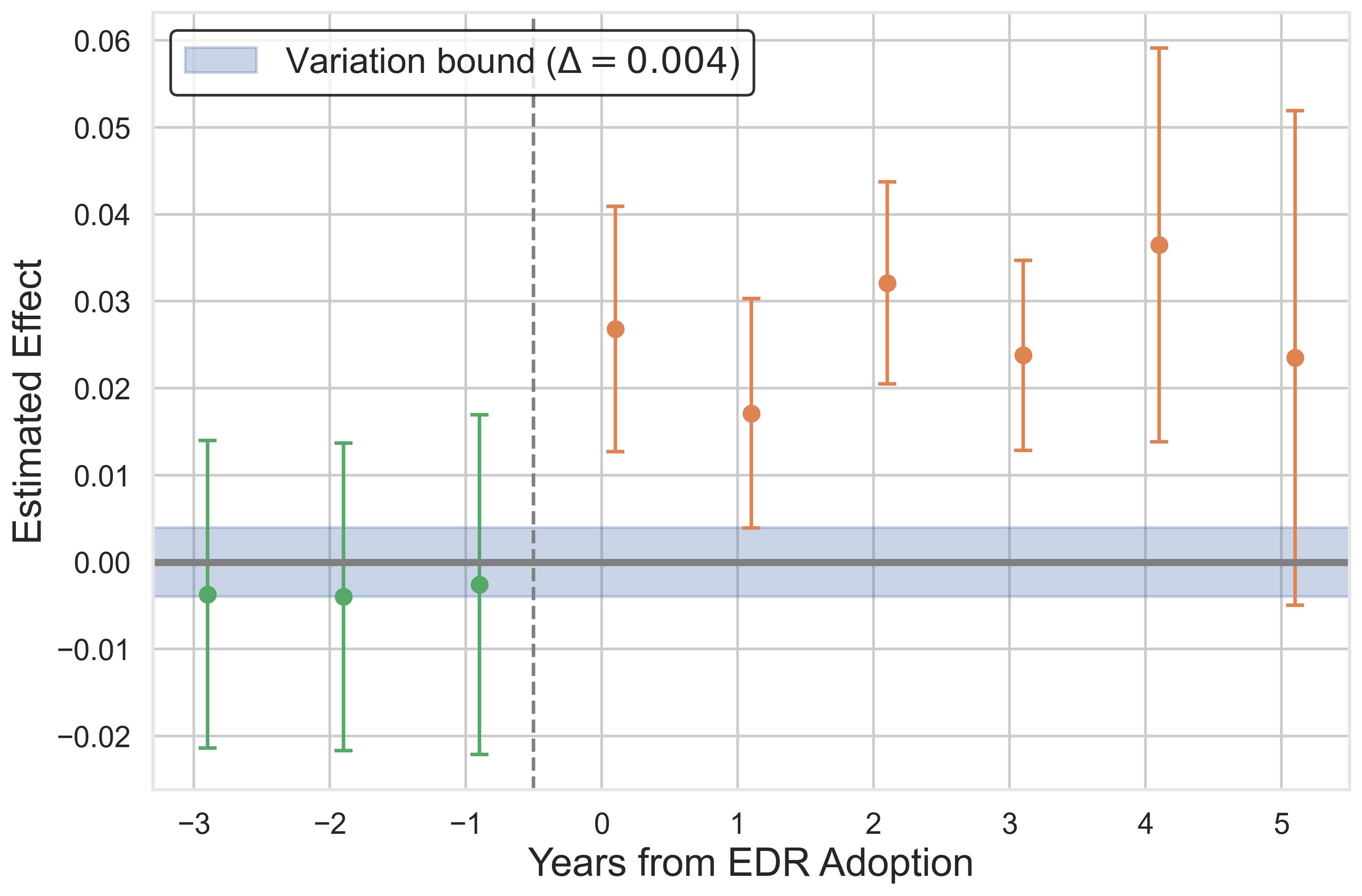}
\caption{) Employment (dummy)}
\label{fig:hd_work_dummy}
\end{subfigure}
\caption*{\footnotesize \textit{Note}:  a) Controls :  2nd order polynomials of marriage duration, husband's age, wife's age,  N. of children and an indicator for education    beyond high school. In case of BJS same variables are used as fixed effects. b) Confidence intervals have a 95\% coverage.c) Families observed for fewer than 5 years post-treatment are excluded from the treatment group to avoid attrition bias.They can, however, serve as control group. d) Variation Bound uses the higher in absolute value coefficient from pre-treatment period as $\Delta$}

\end{figure}

One might argue that the observed effect could be attributed to a shift in bargaining power within the family that diminished women's bargaining power, resulting in reduced leisure time and an increased labor supply for wives. However, Figure~\ref{fig:husband_labor_supply_effect} shows similar results for husbands. This indicates that if EDR did lead to a shift in bargaining power, it wasn't the sole factor driving the positive impact on female labor supply, as a bargaining power shift would likely produce asymmetric effects. Moreover, this result is important not only as a counterargument against the bargaining power shift as an alternative explanation but also as support for the mechanism I propose, which will be discussed later.

Even though Figures \ref{fig:wife_labour_supply_effect} and \ref{fig:husband_labor_supply_effect} demonstrate a significant positive effect of EDR on the labor supply of both spouses, it is important to discuss the validity of these results. For this reason, I provide arguments supporting the validity of the parallel trend assumption.

\begin{enumerate}
    \item \textbf{Unconditional Parallel Trend Assumption}. As previously mentioned, \textcite{borusyak2021revisiting} proposes UPT as a main one. Figure \ref{fig:wife_labour_supply_effect} and Figure \ref{fig:husband_labor_supply_effect} test this assumption for periods prior to EDR adoption. As observed in Figure \ref{fig:wife_labour_supply_effect}, the coefficients prior to EDR adoption (green) deviate from zero, suggesting that the assumption might not hold for wives. However, for husbands in periods prior to EDR adoption (as seen in Figure \ref{fig:husband_labor_supply_effect}), this assumption cannot be rejected. As discussed earlier, this assumption might be too strong, so I proceed with considering weaker ones.

    \item \textbf{Conditional Parallel Trend Assumption.} Given that BJS produces unbiased estimates even under the conditional parallel trend assumption, I use the test employed by CS for periods before EDR adoption to check for significant differences in trends conditioned on covariates. Figure \ref{fig:CPT wife_labour } and Figure \ref{fig:CPT husband_labour } indicate that I cannot reject the assumption since all the $95$\% confidence intervals cover $0$. It should be noted though that, for Figure \ref{fig:CPT wife_labour }b, while two coefficients are insignificant, the reason for non-rejection could be the presence of high standard errors.  However, in general in must be concluded that  \textbf{CPT cannot be rejected} for pre-treatment period

    \item \textbf{Bounded Variation Assumption.} As noted by \textcite{manski2018right} and further generalized by \textcite{rambachan2023more}, even the conditional parallel trend assumption might be too strong. The authors propose an assumption that can be summarized as "the control and treatment groups do not have ``too" different trends in the post-treatment period if the treated group was not actually treated". Following a similar approach to \textcite{manski2018right}, I assume that the trend is bounded by $\Delta$, equal to the largest absolute deviation from zero prior to EDR adoption. 

    The test for this assumption is illustrated in Figure \ref{fig:wife_labour_supply_effect} and Figure \ref{fig:husband_labor_supply_effect} by the shaded area. It is evident that even when allowing for significant trend deviation post-treatment, BJS still produces a significant effect. For example, in later periods, the effect of EDR on wife's weekly hours is at least 2 hours and on the employment rate at least 4 p.p \footnote{Difference between the coefficient and upper bound of shaded area.}. This implies that, even by very conservative estimates, 20\% of the female labor supply increase from 1950 to 2019 can be attributed to EDR.
    
\end{enumerate}

As can be summarized, the parallel trend assumption can only be rejected for its strongest version - UPT - and only for women. In all other assumptions and outcomes, it cannot be rejected
\subsection{Aggregated Effect Over the First 5 Periods after EDR Adoption. }

\begin{table}[h!]
\centering
\caption{ATT aggregated for 5 years after adoption.}
\label{table:ATT_5_years}
\begin{tabular}{ll|cc}
\toprule 
\textbf{Spouse} & \textbf{Variable} & \textbf{BJS (Main)} & \textbf{CS (Secondary)} \\
\toprule 
Husband & Weekly hours & $1.75^{***}$ & $2.17^{**}$ \\
 & & $(0.472)$ & $(0.88)$ \\
 & Employment (binary) & $0.026^{***}$ & $0.038^{***}$ \\
 & & $(0.006)$ & $(0.01)$ \\

Wife & Weekly hours & $2.17^{***}$ & $1.51^{**}$ \\
 & & $(0.482)$ & $(0.567)$ \\
 & Employment (binary) & $0.054^{***}$ & $0.027$ \\
 & & $(0.011)$ & $(0.034)$ \\

Household & Assets (logs) & $0.17^{**}$ & -\\
 &  & $(0.068)$ &  \\

\hline
\hline
\end{tabular}
\begin{center}
\textit{Note:} * $p<0.1$; ** $p<0.05$; *** $p<0.01$.\\
\footnotesize a) Controls included: 2nd order polynomials of marriage duration, husband's age, wife's age, N. of children, and an indicator for education beyond high school. In the case of BJS, the same variables are used as fixed effects. b) Families observed for fewer than 5 years post-treatment are excluded from the treatment group to avoid attrition bias.\footnote{They can, however, serve as control groups.} c) Equation for assets, along with individual and time FE, includes FE for age of wife, duration of marriage, number of children and state.  d) CS is unable to identify the treatment effect for assets due to the significant gaps between waves in the NLSW.
\end{center}
\end{table}

To summarize the findings discussed above, I present the ATT estimates for the first five periods after the adoption of EDR. As can be seen from Table \ref{table:ATT_5_years}, the effect of EDR is positive and significant on the labor supply of both spouses, both in extensive and intensive margins. Moreover, CS can replicate all results produced by BJS; even though the effect on female employment rate is insignificant, it remains positive. Unlike BJS, CS is based on another important assumption, which is the \textit{common support}\footnote{In fact, \textcite{callaway2021difference} posits a weaker assumption, stating that the propensity score must be strictly less than 1}. Figure \ref{fig: Propensity score} demonstrates that the control and treatment groups have significant overlap, and the propensity score is never equal to 1.

\subsection{Mechanism.}

To explain the empirical findings I propose an explanation based on precautionary savings incentives. Divorce is perceived as a form of uncertainty, whether viewed as an exogenous shock or an endogenous choice contingent on other unpredictable factors (``love" shock). Under a TBR, households can adjust the personal share of property of each spouse, by adjusting formal ownership of each unit of property, to ensure consumption smoothness between two marital statuses. The introduction of EDR, however, curtails this flexibility. If couples can’t reach an agreement, the court determines the share of total savings. This reintroduce previously eliminated the uncertainty in consumption level due to risk of divorce, so "prudent" agents respond to it with increased savings. This offsets the difference in utilities between the states of being married and divorced. This holds even if the share is deterministic and known as long as new share chosen by government, is different from the optimal one which would completely eliminate the difference in consumption under different marital statuses in the future. And court involvement just introduces an additional element of uncertainty into the share determination, possibly further enforcing precautionary saving incentives.

In fact, this hypothesis is supported by the findings from Table \ref{table:ATT_5_years}, as EDR also had a positive and significant effect on assets. I cannot produce the same result for savings using CS due to data limitations: the ATT for assets is estimated using NLSW, because PSID does not provide information on assets. This data is not collected annually; therefore, due to these gaps, I cannot use CS as it requires computing the difference between the \textit{specific} post-treatment year and the last year prior to the treatment. For example, I use the last period before adoption ($t=-1$) as the baseline period, so my estimation for CS methods requires computing the first difference $Y_r-Y_{-1}$. However, if for State A, $t=-1$ corresponds to 1970 and NLSW does not survey people in that year, then $Y_r-Y_{-1}$ cannot be calculated. While BJS requires having at least one year from the pre-treatment period instead of using a specific year prior to adoption (e.g., $t=-1$ in my example). Additionally, there is concern regarding the parallel trend assumption for this estimate, which is discussed in more detail in Appendix \ref{sec: Assumption test}. Therefore, I do not rely on it during the model estimation stage.

Moreover, the positive effect on the labor supply of both spouses (see Table \ref{table:ATT_5_years}) can be considered an additional argument supporting the mechanism. This is because the uncertainty affects the entire household, not just individual members, which suggests that it should drive the labor supply of both spouses.

\section{Theoretical Model}\label{sec: model}

To investigate the conditions under which EDR stimulates precautionary savings motives, I have developed a two-period model of a household. In this model, household members make decisions regarding labor, consumption, savings, and the share of property formally owned by each spouse, taking wages and wealth as given during the first period. However, they are unaware of whether they will face a divorce in the second period, which is drawn exogenously with a known probability. In the second period, spouses only make decisions about their consumption levels. In an extension, I demonstrate that the same mechanism prevails even if the divorce is endogenous and spouses are allowed to bargain.

One might argue that the model's design, where household members are unable to work in the second period, excludes other possible mechanisms for addressing increased uncertainty. These could include working more in the second period or enhancing human capital in the first period to ensure higher future income. While this critique is valid, it only suggests additional tools for spouses to respond to increased uncertainty. Nonetheless, the presence of increased uncertainty itself remains the main explanation for my findings.

\subsection{A Model of Household Welfare and Divorce Policy}\label{sec: main model}

Denote household welfare by \( W \), where \( \theta \) represents the weight of the wife's utility in the overall household's welfare, interpreted as the bargaining power of the wife. The utility functions of the wife and husband are denoted by \( U^f \) and \( U^m \), respectively. Welfare is then determined as follows:
\begin{multline}\label{eq: Welfare function}
    W=\theta U^f(c_{1,f},l_{1,f})+(1-\theta)U^m(c_{1,m},l_{1,m}) \\
    + \beta \Big[ (1-\pi) \Big( \theta U^f(c_{2,f},0)+(1-\theta)U^m(c_{2,m},0) \Big) \\
    +\pi E\Big(\theta U^f(c_{2,f,d},0)+(1-\theta)U^m(c_{2,m,d},0) \Big) \Big]
\end{multline}

\noindent where \( c_{t,j} \) is the consumption of spouse \( j \) at time \( t \), \( l_{t,j} \) is the labor supply of spouse \( j \) at time \( t \), \( \pi \) is the probability of the couple getting divorced, and \( \rho \) is the share of property under the wife's name.

Formally, the household solves the following maximization problem:
\begin{align}\label{eq: Maximization Title}
\max_{c_{1,f}, c_{1,m}, l_{1,f}, l_{1,m}, s, \rho} \quad & W \\
\text{s.t.} \quad & c_{1,f} + c_{1,m} + s \leq w_{1,f} l_{1,f} + w_{1,m} l_{1,m} + S_0 \quad \text{(where \( S_0 \) is the initial wealth)} \\
& c_{2,f} + c_{2,m} \leq s \quad \text{if married in second period} \\
& c_{2,f,d} \leq \alpha s \quad \text{if divorced in the second period} \\
& c_{2,m,d} \leq (1-\alpha) s \quad \text{if divorced in the second period} \\
& \alpha = \rho  \quad \text{if TBR} \\
& \alpha \sim F(\alpha)  \quad  \text{if EDR} 
\end{align}

\noindent where \(\alpha\) can be interpreted as the divorce policy. In particular:
\begin{itemize}[label=\textasteriskcentered]
    \item \textbf{TBR:} The share, \(\alpha\), corresponds to the share of property under the wife's name, \(\rho\).
    \item \textbf{EDR:} \(\alpha\) is determined by the court, so I treat it as random. In the main model, spouses cannot agree on a share without court intervention. Nevertheless, subsequent analysis will show that allowing spouses to mutually decide on property share and divorce endogenously has no impact on the main outcomes.
\end{itemize}

\vspace{1em} 
\subsection{EDR and Precautionary Savings}
In this section, I compare TBR to EDR to characterize the conditions under which EDR stimulates the precautionary savings motives of husbands and wives in the first period of the model.  

\subsubsection{Sufficient Condition for EDR to Stimulate Precautionary Savings: General Utility Function}

To simplify the analysis, I focus on the sufficient conditions that must be met for EDR to increase household savings. 

The first finding is that the household prefers to completely smooth consumption under TBR:

\textbf{Proposition 1:}
\textit{Assume that:}
\begin{itemize}
    \item $U^j_c(0,l)=\infty$
    \item $U_c^j(c,l)>0$ - \textit{monotonous preferences}
    \item $U_{cc}^j(c,l)<0$ - \textit{decreasing marginal utility}
\end{itemize} 

\textit{Then the household \textbf{fully} insures against the divorce shock, i.e.,} $c_{2,j}=c_{2,j,d} \quad \forall j \in \{f,m\}$.

The proof is in Appendix \ref{sec: Sufficient condition}.

Since EDR leads to a different allocation, it must not lead to full insurance. 

The second finding is that EDR stimulates precautionary savings motives if weighted average of spouses' the \textit{relative prudence}, $RP(x)=-x\frac{u'''(x)}{u''(x)}$, is higher than $2$, and leads to lower savings than under TBR if it is lower than $2$.

\textbf{Proposition 2: Sufficient Condition for EDR's Effect on Savings Incentives}

\textit{Let \( H \) be defined as:}
\begin{equation}
    H = (1-\alpha) \times RP^f(\alpha s) + \alpha \times RP^m((1-\alpha) s)
\end{equation}
\textit{where}
\begin{equation}
    RP^j(x) = -x\frac{U_{ccc}^j(x,0)}{U_{cc}^j(x,0)} \quad \text{\textit{denotes the relative prudence.}}
\end{equation}
\textit{For EDR to affect savings incentives, the following conditions must be met:}

\begin{enumerate}
    \item \textit{The household's optimal wife's share under TBR, $\alpha^*$, satisfies:}
    \begin{equation}
        \frac{U^f_{cc}(\alpha^* s,0)}{U^m_{cc}((1-\alpha^*) s,0)} = \frac{1-\theta}{\theta}\times \frac{1-\alpha^*}{\alpha^*}
    \end{equation}
    
    \item \textit{Regarding the effect on savings:}
    \begin{itemize}
        \item \textit{If \( H > 2 \), then EDR leads to an increase in savings.}
        \item \textit{If \( H < 2 \), then EDR leads to a decrease in savings.}
        \item \textit{If \( H = 2 \), then EDR has no effect on savings.}
    \end{itemize}
\end{enumerate}

\textit{Here, \( \alpha^* \) is the optimal share of the wife in the case of divorce, which the household would choose under TBR. \footnote{Do not confuse $\alpha$ with $\rho$. While $\rho$ simply represents the share of property owned by the wife, in my model it equals the share she receives in the event of a divorce under TBR. For instance, an actual policy could guarantee a minimal share each spouse can receive under TBR, such as $\alpha = \max(\rho, \bar{\alpha})$, where $\bar{\alpha}$ is the minimal share specified by the policy.}}

The proof is in Appendix \ref{sec: Sufficient condition}.

The condition (2) from Proposition 2 can be considered a generalization of the one derived by \textcite{vergara2017precautionary} for a single agent. Instead of individual $RP$, precautionary saving incentives for a household are determined by the weighted average of the spouses' $RPs$. The condition (1) is needed to ensure that \textit{any} deviation of $\alpha$ from $\alpha^*$ leads to an increase (or decrease) in savings if the FOC for savings, $s$, is convex (or concave) in $\alpha$.

I now provide a condition under which EDR increases the labor supply through precautionary savings.

\textbf{Proposition 3:} 
\textit{The labor supply of spouse $j$ will increase (or decrease) with higher (or lower) precautionary savings incentives if and only if:} 
\begin{enumerate}
    \item \textit{Precautionary savings result in higher consumption in the first period, $c_{1,j}$.}

    \textit{or} 

    \item \textit{Precautionary savings lead to decreased $c_{1,j}$ and} 
    \begin{equation}
        -\frac{U^j_{lc}(c_{1,j},l_j)}{U^j_{cc}(c_{1,j},l_j)} < w_j
    \end{equation}
\end{enumerate}

The proof can be found in Appendix \ref{sec: Sufficient condition}.

\textbf{Separable Utility:} $U^j(c,l)=u(c)-h(l)$. Given this form, $U^j_{lc}(c_{1,j},l_j)=0$ for all values of $c_{1,j}$ and $l_j$, and since $w_j>0$, any increase (or decrease) in precautionary savings will invariably lead to a corresponding increase (or decrease) in the labor supply of spouse $j$.

\subsubsection{Special Case: CRRA.}\label{sec:CRRA condition main}

In this section, the focus is on the special case where the utility function exhibits Constant Relative Risk Aversion (CRRA). Specifically, I will derive the \textit{necessary} and \textit{sufficient} condition for EDR to increase or decrease the labor supply of each spouse.

Let's assume that the utility function for each spouse is separable:

\begin{equation}
U^j(c_j,l_j) = \frac{c_j^{1-\gamma}}{1-\gamma} - f^j(l_j)
\end{equation}

where $f^j(0) = 0$. Both spouses share the same coefficient of relative risk-aversion, denoted by $\gamma$.

\textbf{Corollary 1: Precautionary Savings Condition.}
 \textit{Upon inspection, it becomes apparent that condition $1$ from Proposition 2 is always satisfied by CRRA utility. This implies that any deviation from $\alpha = \alpha^*$ will result in an increase in savings when $H = 1 + \gamma > 2 \quad (\gamma>1)$ or a decrease when $H = 1 + \gamma < 2 \quad (\gamma<1)$.}

\textbf{Proposition 4:} \textit{Given the utility function \( U^j(c_j,l_j) = \frac{c_j^{1-\gamma}}{1-\gamma} - f^j(l_j) \), the sign of the difference in labor supply of spouse \( j \) between EDR and TBR is determined by \( \gamma \) as follows:}

\[
l_j^{EDR} - l_j^{TBR}
 =
\begin{cases} 
    > 0 & \Leftrightarrow  \gamma > 1 \\
    < 0 & \Leftrightarrow  \gamma < 1 \\
    = 0 & \Leftrightarrow  \gamma = 1
\end{cases}
\]

\textit{Here, \( l_j^{EDR} \) and \( l_j^{TBR} \) represent the labor supply of spouse \( j \) under the EDR and TBR regimes, respectively.}

\textbf{Proof: }The proof leverages the conditions outlined in Proposition 2, which generally holds as a sufficient condition only. However, for the CRRA utility of consumption, it stands as an ``if and only if" condition. Furthermore, the condition stated in Proposition 3 always holds for \( U^j(c_j,l_j) = \frac{c_j^{1-\gamma}}{1-\gamma} - f^j(l_j) \) due to its separable nature. This proves the Proposition 4.\\

\par
Proposition 4 plays a key role throughout this paper. It posits that precautionary savings are the \textbf{sole} mechanism impacting the sign of the treatment effect of EDR on labor supply. This is underscored by the fact that the statement in the proposition stands as an ``if and only if" condition.

\section{Econometric Implementation and Identification}\label{sec: Econometric and identification}
I now estimate the model so that the mechanism discussed in the previous section can be quantified. To better match the data, I specify the model as follows:
\begin{enumerate}
\item Rather than the general utility function $U^j$ mentioned in Section \ref{sec: main model}, I adopt a parametric functional form: \textbf{CRRA}.
\item To rationalize the variance in household choices that are similar based on observed characteristics, I introduce \textbf{unobserved shocks}.

\item To help the model fit labor supply for different age groups, I introduce \textbf{heterogeneity} in wages and the disutility of work, particularly in relation to the wife's age.
\end{enumerate}
Together, these three modifications add realism to the model while also facilitating its mapping to the empirical moments.
 
\subsection{Econometric Specification}

\paragraph{Preference.}
Agents derive positive utility from consumption and experience negative utility (disutility) from labor. The utility is described by individual consumption, \(c_{t,j}\), and labor supply, \(l_{t,j}\):

\begin{equation}
    U^j(c_{t,j},l_{t,j})=\frac{c_{t,j}^{1-\gamma}}{1-\gamma}-\phi_j l_{t,j}
\end{equation}

Furthermore, I specify $\phi^j$ as

\begin{equation}
    \log(\phi_j)=\phi^j_0+\phi^j_1 \cdot t+\phi^j_2 \cdot t^2 
\end{equation}

Here, \(t\) is the age of the wife. When \(j=m\) (husband), the wife's age is used as a proxy for the husband's age, allowing \(\phi_j\) to vary with age so that the model can rationalize the decrease in labor supply with age.

\paragraph{Choices.}
 Consumption ($c$) and savings ($s$) are continuous and are computed from the first-order condition. Labor supply ($l$) is discretized on an equally spaced grid of size 8, ranging from 0 to 40 for each spouses.

\paragraph{Wages and Initial Wealth.}
Wages are functions of the wife's age, serving as a proxy for work experience for both spouses, and are also influenced by wage shocks:

\begin{equation}
     \log(w_{j})=\bar{w}^{j}+\delta_1^j t+\delta_2^j t^2+\varepsilon_{j}
\end{equation}

Wage shocks of both spouses are assumed to be jointly normally distributed.

\begin{equation}
    (\varepsilon_{f},\varepsilon_{m}) \sim N(0, \Sigma)
\end{equation}

While the wealth shock is also normally distributed, I assume that it is independent of the wage shock. This might be a strong assumption; however, it is made due to the fact that the wealth moment comes from a different dataset, and I do not have a moment correlating wage shocks with wealth. Therefore, the correlation with wage shocks may not be point-identified. It is also important to note that I use the difference $s - S_0$ to match the moment instead of matching savings, $s$, and initial wealth, $S_0$, separately.

\begin{equation}
    S_0 \sim N(\mu_S, \sigma_S)
\end{equation}

\paragraph{Property Share under Equitable Distribution.}
The share \(\alpha\) is assumed to follow a Beta distribution due to inherent uncertainties faced by households, its flexibility, and its ability to accommodate values between 0 and 1. Insights from \textcite{woodhouse2006divorce} suggest that secondary earners typically secure between one-third and one-half of total savings. I therefore restrict the expected share of wife to be between $\frac{1}{3}$ and $\frac{1}{2}$.

\paragraph{Preset Parameters of the Model.}
Certain parameters, not pivotal for the project or lacking point identification, are predetermined based on existing literature (Table \ref{Table: preset-parameters}).

\begin{table}[h!]
    \centering
    \caption{Preset Parameters}\label{Table: preset-parameters}
    \begin{tabular}{lcl}
    \toprule
    Parameters & Values & Reference \\
    \midrule
    Discount factor, \(\beta\) & 0.98 & \\
    Probability of divorce, \(\pi\) & 0.2 &  \textcite{schweizer2020divorce}\footnote{Approximately 20\% of women who have ever been married are divorced or separated.} \\
    Bargaining power of wife, \(\theta\) & 0.3 & \textcite{voena2015yours} \\
    \bottomrule
   
    \end{tabular}
    
     \footnotesize \textit{Note:} I use the probability of getting ever divorced for existing\\  families in 1970s as probability of divorce, $\pi$.
\end{table}

\subsection{Identification}

Given the standard nature of this model, our focus is on the key identification challenges: 

\begin{enumerate}
    \item Identification of the wage function.
    \item Determining the relative risk aversion coefficient, \(\gamma\).
    \item Estimating the variance of the wife's share, \(\sigma_\alpha^2\).
\end{enumerate}

\textbf{Wage function.} Reduced form results indicate that policy impacts labor supply. Given the challenges in conceiving a situation where such policies would directly affect individual productivity, it's more plausible that they predominantly influence the household decision-making process. This points to the law or redistribution regime as a variable affecting labor participation decisions while being excluded from the wage equation, which facilitates the identification of wage equation parameters.

\textbf{Relative risk aversion, \(\gamma\).} The relative risk aversion is pivotal as it directly affects the sign of the EDR treatment effect. The analysis (detailed in Appendix \ref{sec: CRRA condition}) suggests that the effect varies with \(\gamma\): negative for \(\gamma<1\), neutral for \(\gamma=1\), and positive for \(\gamma>1\).  Moreover, as it will demonstrated later (Figure \ref{fig:ATT gamma: EQ-ttl}) treatment effect on labor is an increasing function of  $\gamma$ so as long as the rest of the parameters are identified by moments other then treatment estimates from Table \ref{table:ATT_5_years} I can point identify the $\gamma$.

.

\textbf{Distribution of wife's share of assets upon divorce, \(\alpha\).} The distribution of \(\alpha\), denoted as \(B(a, b)\), is characterized by two parameters: \(a\) and \(b\). Together with \(\gamma\), they are the only parameters that influence the treatment effect estimates within my model. Concurrently, I have \(4\) treatment estimates (See Table \ref{table:ATT_5_years}). As long as at least \(3\) moments of those \(4\) are linearly independent, identification must be achievable. The treatment effect on labor supply for both husbands cannot be fully correlated. However, their treatment effect on employment status and working hours is correlated, since one outcome variable is a function of another. Yet, this relationship is not linear. Some individuals might work both prior and post-EDR, but they could have engaged in longer weekly hours under EDR. This allows me to assert that all three parameters \(\gamma\), \(a\), and \(b\) must be \textbf{point-identified.}

\textbf{Identification of other parameters.} The identification of the remaining parameters is achieved by the set of moments I employ. The disutility parameters, $\phi^j_0, \phi^j_1,\phi^j_2$, are identified by the labor supply of each spouse, $j$, conditional on the age of the wife, $t$. The distribution of the wage shock is identified by the variance and correlation of accepted wages.Identification of  variance, $\sigma_S$, and mean, $\mu_S$, of initial wealth, $S_0$ is achieved by matching with  the mean and variance of the change in assets (savings), denoted as $\overline{assets_t-assets_{t-1}}$ . The identification of these parameters ensures that having the four treatment estimates is enough for the over-identification of the three crucial parameters: $\gamma, a,$ and $b$. In addition to the aforementioned moments, I also match  several others, such as the correlation between labor supply decisions and  correlation between wage and labor supply choices, among others.

\subsection{Implementation.}

The estimation process unfolds in several pivotal steps: Initially, the model with EDR is simulated to compute various moments such as average labor supply, savings, and the correlation between labor, etc. (\textit{Step a}). Subsequently, for the same people, a simulation under the title-based regime is conducted to compute their labor supply choices (\textit{Step b}). The ensuing phase involves the computation of the squared difference between the moments determined in \textit{Step a} and their corresponding empirical moments (\textit{Step c}). Hereafter, the squared difference between the difference in labor supply produced by the EDR model in \textit{Step a} and by the title-based model in \textit{Step b} for the same people, and the ATT  estimates ,$\Bar{\tau}$ \footnote{This estimate corresponds to the one in Table \ref{table:ATT_5_years}}, is calculated as:
\begin{equation}
\frac{\Big (E(l_{\text{EDR}}) - E(l_{\text{Title Based}}) - \Bar{\tau}\Big )^2 }{\Bar{V}(\Bar{\tau})}
\end{equation}
(\textit{Step d}). The final step is to minimize the summ of the quantities from \textit{Step c} and \textit{Step d} to get optimal parameter estimates.

\subsection{Estimates of Model Parameters.}

 Before proceeding with the model estimation, I estimate the wage equations using Heckman correction. This approach helps to reduce the number of parameters explicitly estimated within the model. Table \ref{Table: Wage equation} shows the results of this exercise.
 \begin{table}[h!]
\centering
\caption{Wage Equation Parameters} \label{Table: Wage equation}
\begin{tabular}{lcc}
\toprule
& \multicolumn{1}{c}{Log Wife Wage} & \multicolumn{1}{c}{Log Head Wage} \\
\midrule
Constant & $0.23$ $(0.098)$ & $0.24$ $(0.095)$ \\
Age of Wife & $0.01$ $(0.004)$& $0.026$ $(0.004)$ \\
Age of Wife Squared & $-0.00018$ $(0.00006)$ &  $-0.00028$ $(0.00006)$ \\
\bottomrule
\end{tabular}

\smallskip
\footnotesize{\textit{Note:} Standard errors are in parentheses. The sample is restricted to observations under Title-based \\regimes or to the first $5$ periods under equitable distribution regimes. All states are mutual consent states.}
\end{table}

Table \ref{table: estimation results} presents the parameters estimated internally. As can be seen from the table, the relative risk-aversion coefficient, \(\gamma\), is $2.09$, which is notably higher compared to the value of $1.5$ used by \textcite{attanasio2008explaining} and \textcite{voena2015yours}. It is worth noting that this value is not explicitly estimated by those authors; they simply choose it to be equal to $1.5$ as it is consistent with the elasticity of intertemporal substitution estimated by \textcite{attanasio1995consumption}. This deviation could be attributed to the inherent limitations of  a two-period model in this study. For instance, in scenarios where individuals might be inclined to work despite low wages and high disutility, anticipating substantial returns from accumulated human capital, a model constrained to agents being able to work only in the first period is unable to adequately represent such long-term strategic behaviors. Consequently, behaviors rationalized as strategic investments in human capital in a dynamic setting are interpreted as  high risk-aversion (higher \(\gamma\)) in my model, given its inability to encompass multiple periods of work and experience accumulation.

 The distribution of $\alpha$ described by   expected value \(E(\alpha) = \frac{a}{a+b} = 0.49\) and a variance \(Var(\alpha) = \frac{ab}{(a+b)^2 (a+b+1)} = 0.07\), thereby yielding a standard deviation of approximately \(0.26\). This implicates a high level of uncertainty regarding the expected share under EDR. This could be explained by the fact that I focus on the first 5 years after EDR adoption, so agents might have little information about the actual implementation of the EDR, and their beliefs could be different from the bounds provided by \textcite{woodhouse2006divorce}. However, in my model I  assume that agents have "rational" expectation - their belief about distribution of $\alpha$ (distribution I estimate) is the actual distribution. I do this my putting the restriction on $E(\alpha)$ to be within bounds from \textcite{woodhouse2006divorce}.

Table \ref{tab:model-fit} demonstrates the model’s ability to accurately replicate the key moments from the data, despite a less precise match for the difference in savings, a discrepancy attributed to the small sample size of the NSLW (N=265). The reason behind having a small sample size is that I had to drop many states to focus only on those with Mutual Consent, and given that the first difference in savings requires the agents to be observed at least two times in the dataset, I had to drop those which are observed only once. However, Tables \ref{tab:model-fit} and \ref{Table: Treatment match} collectively confirm the model's ability to closely replicate both observed and counterfactual scenarios (as I consider the treatment effect estimate as a "bridge" connecting data to the counterfactual scenario).

\begin{table}[h!] 
    \centering
     \caption{Estimation results}\label{table: estimation results}
   \begin{tabular}{lc|cc}
\toprule
 \textbf{Parameter} & \textbf{Value (SE)} & \textbf{Parameter}& \textbf{Value (SE)} \\

     \midrule
              $\gamma$ &   2.09 (0.043) &      $\sigma_{wf}$ &   1.50 (0.084)  \\
$\phi_f^0$ &  -5.65 (0.47) &            $\mu_S$ &  17.90 (33.37) \\
$\phi_f^1$ &  -0.26 (0.11) &     $\sigma_{w_m}$ &   1.23 (0.07) \\
$\phi_f^2$ & 0.0081 (0.0079) &       $\sigma_{S}$ &   1.28 (0.11) \\
$\phi_m^0$ &  -7.90  (0.48)& $\sigma_{w_f,w_m}$ &   0.70 (0.3) \\
$\phi_m^1$ &  -0.10 (0.039)&        $E(\alpha)$ &   0.49 (0.005) \\
$\phi_m^2$ & 0.0048  (0.026)&        $V(\alpha)$ &   0.07( 0.0008)\\
\bottomrule
\multicolumn{4}{l}{\textit{Note:} \footnotesize The standard errors (SE) are estimated using 170 bootstrap draws.} \\
\end{tabular}

    \label{tab:my_label}
\end{table}

\begin{table}[h!]
    \centering
    \caption{Model Fit to Data}\label{tab:model-fit}
    \begin{tabular}{l|cc}
        \toprule

         \textbf{Moment}  &\textbf{Model}& \textbf{Data}\\
        \midrule
     
  $E(l_f)$ &         22.90 &  20.54 (0.24) \\
   $E(l_m)$ &         32.29 &  35.71 (0.49) \\
 $P(l_f>0)$ &          0.65 &   0.69 (0.01) \\
 $P(l_m>0)$ &          0.88 &   0.88 (0.012) \\
 $E(s-S_0)$ &         99.62 & 46.64 (13.04) \\

        \bottomrule
    \end{tabular}
\end{table}

\begin{table}[h!]
    \centering
    \caption{Ability of model to replicate the treatment effect.}\label{Table: Treatment match}
    \begin{tabular}{c|c|c}
    \toprule
       \textbf{Variable}  &\textbf{Model}  & \textbf{Data} \\
       \midrule
             $E(l_f|D=1)-E(l_f|D=0)$ &  2.445 &  2.17(0.482) \\
    $E(l_m|D=1)-E(l_m|D=0)$ &  1.582 &  1.75(0.472) \\
$P(l_f>0|D=1)-P(l_f>0|D=0)$ &  0.052 & 0.054(0.011) \\
$P(l_m>0|D=1)-P(l_m>0|D=0)$ &  0.025 & 0.026(0.006) \\
            \bottomrule
            \bottomrule
            \multicolumn{3}{l}{\textit{Note:} \footnotesize Treatment effect estimates are taken from Table \ref{table:ATT_5_years}.} \\
    \end{tabular}
    
    \label{tab:my_label-2}
\end{table}

. 

\section{Counterfactual Simulations}\label{sec: Counteractual}
\subsection{Precautionary Savings and Robustness Checks}\label{subsec: ATT-gamma}

The central hypothesis posits that the treatment effect observed in Table \ref{table:ATT_5_years} is driven by precautionary savings due to changes in property distribution law. Moreover in section \ref{sec: CRRA condition} it was shown that within my model precautionary savings is the \textbf{only} mechanism defining the sign of the effect of EDR on household labor. So the goal of this section is two-fold: a) quantify the effect of $\gamma$ on difference in labor supply under EDR and TBR (treatment) b) show that significant extension of model does not change results produced by main) model. 
\subsubsection{Quantifying the Role of $\gamma$ for Labor Supply and Savings.}
To understand the effect of $\gamma$ on findings from Table \ref{table:ATT_5_years}, I plot $E\Big[l_j(EDR,\gamma)-l_j(TBR,\gamma)\Big]$ and $E\Big[s(EDR,\gamma)-s(TBR,\gamma)\Big]$. This relationship is not feasible for direct estimation from data, as $\gamma$ does not vary. However, the estimated model allows for the simulation of different values of $\gamma$, thereby enabling the understanding of the actual heterogeneity of treatment along this dimension.

\begin{figure}[h!]
    \centering
    \caption{Treatment effect size vs  risk-aversion,$\gamma$.}
    \label{fig:ATT gamma: EQ-ttl} 
    
    \begin{subfigure}{0.3\textwidth}
        \caption{ ) Wife's Weekly  Hours}
        \includegraphics[width=\linewidth]{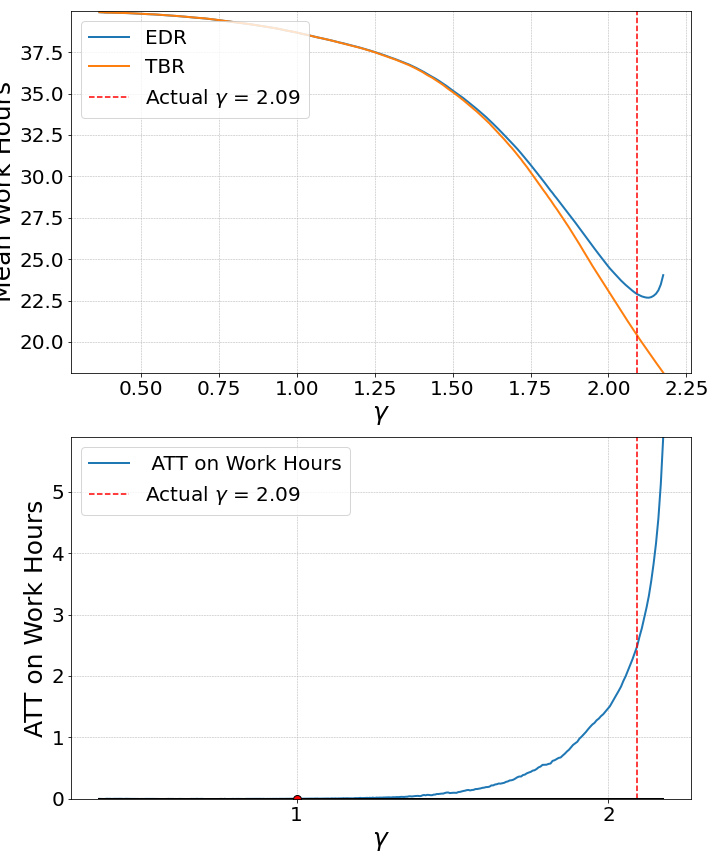}
        \label{fig:ATT gamma: EQ-ttl a}
    \end{subfigure}\hfill
    \begin{subfigure}{0.3\textwidth}
        \caption{ ) Husband's Weekly  Hours}
        \includegraphics[width=\linewidth]{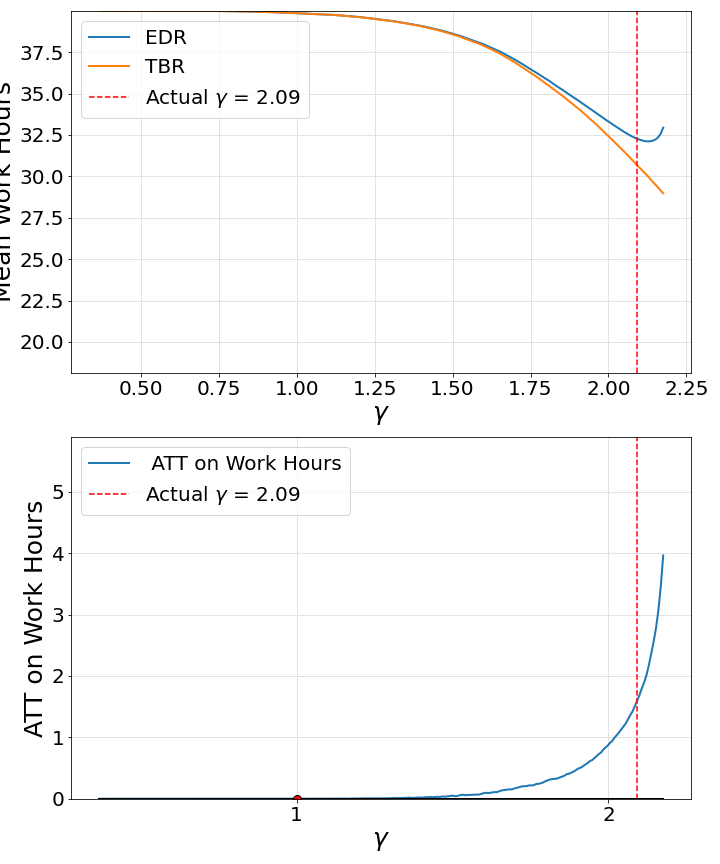}
        \label{fig:ATT gamma: EQ-ttl b}
    \end{subfigure}\hfill
    \begin{subfigure}{0.3\textwidth}
        \caption{) Weekly change in savings }
        \includegraphics[width=\linewidth]{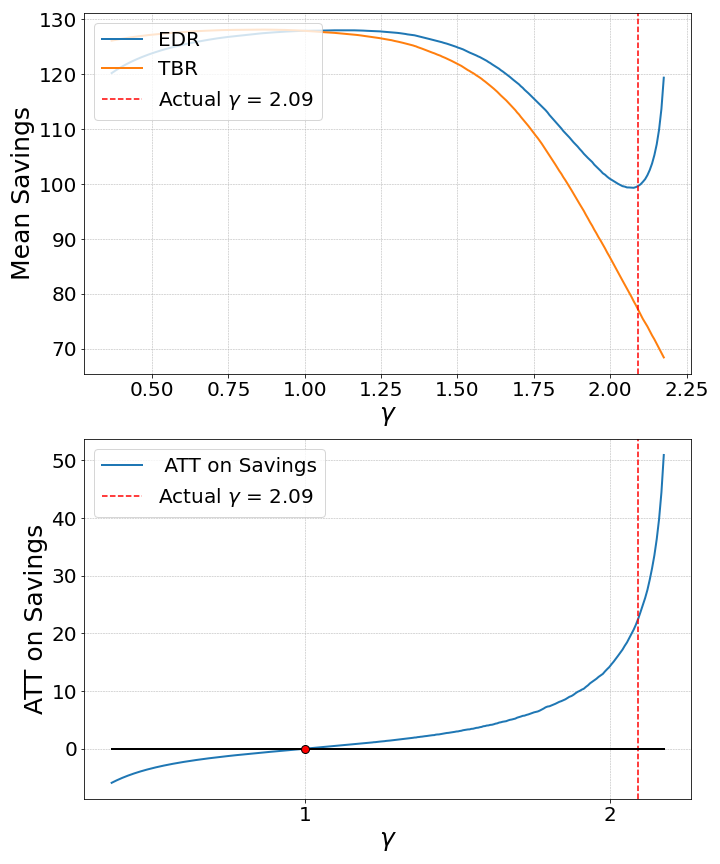}
        \label{fig:ATT gamma: EQ-ttl c}
    \end{subfigure}
       \caption*{\footnotesize \textit{Note}: a) The upper plots show the change in the mean of labor supply and savings with the coefficient of relative risk-aversion, $\gamma$. b) The bottom plots illustrate how the treatment effect (ATT) on labor supply and savings varies with $\gamma$. c) ATT is used to denote the difference in labor supply, $E\Big[l_j(EDR,\gamma) - l_j(TBR,\gamma)\Big]$, and in savings,$E\Big[s(EDR,\gamma) - s(TBR,\gamma)\Big]$, under different regimes.}
\end{figure}

Figure \ref{fig:ATT gamma: EQ-ttl} plots levels (top plot) of labor supply and savings and the treatment effect (bottom plot), which is the difference in outcomes between EDR and TBR. As can be seen across choices of $\gamma$, both spouses are inclined to supply the maximum amount of labor for relatively low values of $\gamma$, resulting in no change in labor supply, making $E\Big[l_j(EDR,\gamma)-l_j(TBR,\gamma)\Big]=0$ (bottom plot). However, with higher $\gamma$, husbands and wives start to work less. This decrease is explained by the effect of $\gamma$ on the marginal utility of consumption, $u_c'=\frac{1}{c^\gamma}$, which decreases with increasing $\gamma$ for all $c>1$. Consumption becomes less attractive relative to leisure, leading them to work less. However, as shown in Section \ref{sec:CRRA condition main}, higher $\gamma$ increases precautionary motives under EDR, which pushes labor supply up, making the decrease in labor supply less steep with increasing $\gamma$. Under TBR, $\gamma$ does not affect the precautionary savings motive due to the lack of uncertainty – people completely insure themselves against divorce risk by smoothing consumption across two marital statuses. This results in a steeper decrease in labor supply under TBR. Consequently, the treatment effect of EDR on labor supply decreases; the same logic can be applied to savings.

From Figure \ref{fig:ATT gamma: EQ-ttl}, it is evident that the point $\gamma=1$ plays a pivotal role. At this value, the treatment effect for savings switches its sign, which aligns with the results derived from Proposition 2 for the CRRA utility function. Notably, the estimates suggest that spouses tend to supply the maximum amount of labor for values of $\gamma$ that are less than 1 under both regimes, resulting in no noticeable difference. While I have explicitly demonstrated that the sign must flip at the value of $\gamma=1$ for labor supply, I will further illustrate this graphically in the next section.

While the argumentation provided so far supports the presence of precautionary savings incentives, it is essential to underscore that a key assumption in my model is \textit{exogenous divorce}. Currently, my model does not allow for the possibility of spouses mutually bargaining over a share at the moment they opt for divorce, making my model a unitary one. In the upcoming section, I will present robustness checks by relaxing these assumptions.
 
\subsubsection{Robustness check.}
While the primary model of this paper assumes that divorce is an exogenous event and does not allow for the possibility of spouses mutually bargaining over a share, these assumptions might be considered too strong. This section aims to ensure that my main conclusions remain valid even when these assumptions are significantly relaxed.

The key finding of this study is the pivotal role of precautionary savings motives. As demonstrated, the sign of the labor supply effect is firmly anchored by relative prudence, defined as \( RP = 1 + \gamma \). Proposition 4 states that the impact of EDR on labor supply and savings switches its sign at \( RP = 2 \) or equivalently \( \gamma = 1 \). However, simulations derived from the parameters in Table \ref{table: estimation results} consistently yield labor supply values around \( \gamma = 1 \), as showcased in Figure \ref{fig:ATT gamma: EQ-ttl}. This phenomenon is attributed to labor supply being constraint by an upper bound, $l_{j}\leq40$ .\footnote{Consider a situation where, without constraints, the optimal labor supply under EDR is 50 hours, and under TBR, it is 45 hours. This would typically lead to a treatment effect of 5 hours. However, if there's an upper limit of 40 hours, the labor supply for both scenarios is capped at this limit, effectively reducing the treatment effect to zero.}

To counteract this, I've adjusted the disutility of labor, increasing its magnitude, thereby inducing agents to offer less labor. For ease of reference, these agents are termed as ``lazy''. Their disutility parameter is derived as \( \phi^0_{j,lazy} = \phi^0_{j} + \Delta_j \), where \( \phi^0_{j} \) is sourced from Table \ref{table: estimation results}. The magnitude of \( \Delta_j \) is manually chosen for achieving a non-binding labor supply around \( \gamma = 1 \) and might differ across robustness check scenarios.

Before diving into the robustness checks, a simulation reinforcing Proposition 4 is presented. While seemingly reiterative, the proposition is already proven, this exercise serves as a foundational reference for the subsequent robustness checks.

\begin{figure}[h!]
    \centering
    \caption{Treatment effect size vs  risk-aversion for ``lazy" agents,$\gamma$.}
    \label{fig:ATT gamma: EQ-ttl lazy} 
    
    \begin{subfigure}{0.3\textwidth}
        \caption{ ) Wife's Weekly  Hours}
        \includegraphics[width=\linewidth]{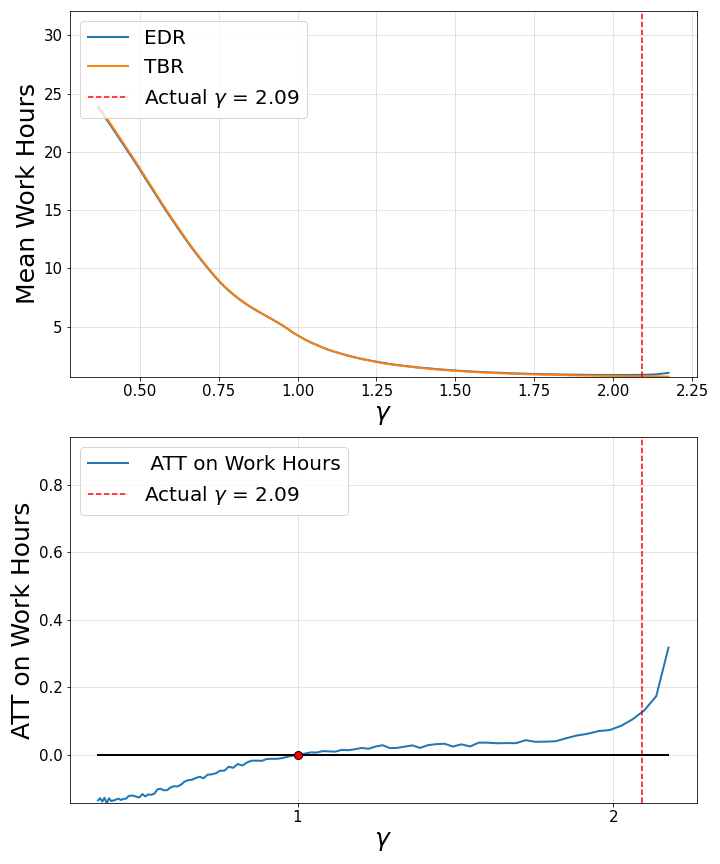}
        \label{fig:ATT gamma: EQ-ttl a-2}
    \end{subfigure}\hfill
    \begin{subfigure}{0.3\textwidth}
        \caption{ ) Husband's Weekly  Hours}
        \includegraphics[width=\linewidth]{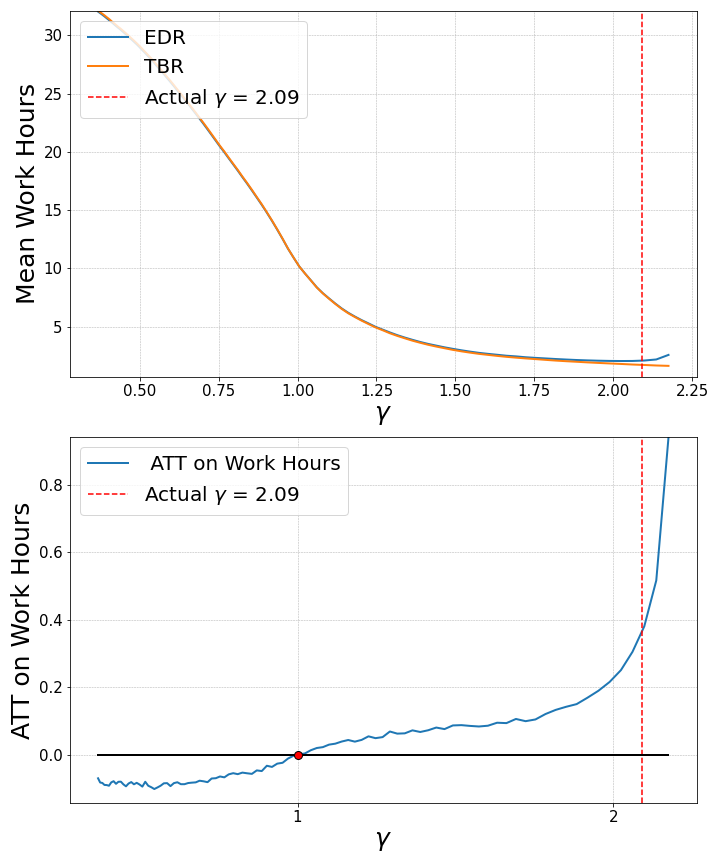}
        \label{fig:ATT gamma: EQ-ttl b-2}
    \end{subfigure}\hfill
    \begin{subfigure}{0.3\textwidth}
        \caption{) Weekly change in savings }
        \includegraphics[width=\linewidth]{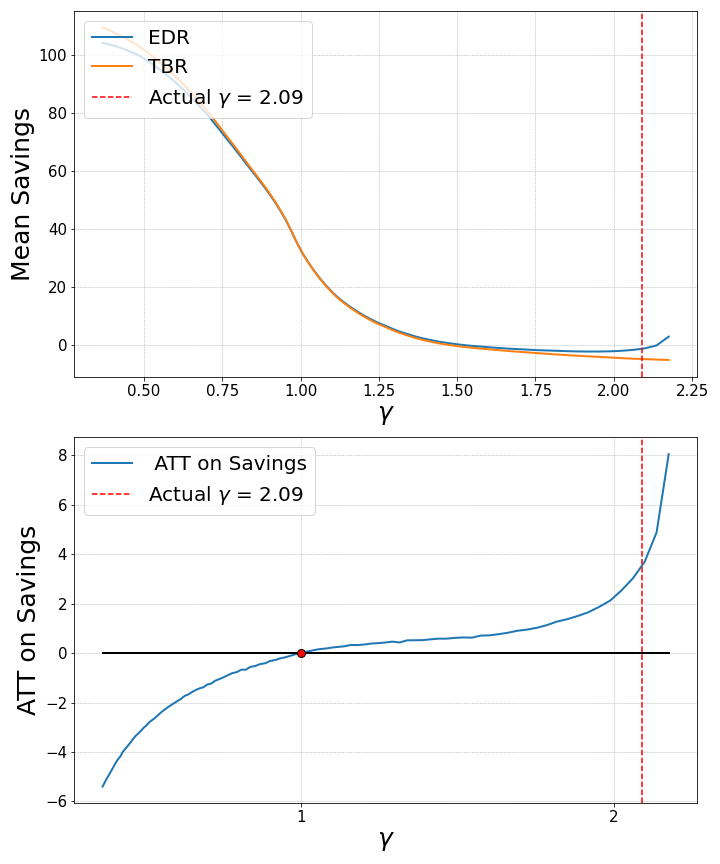}
        \label{fig:ATT gamma: EQ-ttl c-2}
    \end{subfigure}
    \caption*{\footnotesize \textit{Note}: a) The upper plots show the change in the mean of labor supply and savings with the coefficient of relative risk-aversion, $\gamma$. b) The bottom plots illustrate how the treatment effect (ATT) on labor supply and savings varies with $\gamma$. c) ATT is used to denote the difference in labor supply, $E\Big[l_j(EDR,\gamma) - l_j(TBR,\gamma)\Big]$, and in savings,$E\Big[s(EDR,\gamma) - s(TBR,\gamma)\Big]$, under different regimes.}
\end{figure}

\textbf{Baseline Model Results for ``Lazy'' Agents:} Parameters from Table \ref{table: estimation results} form the basis of this simulation, with a modified \( \phi^0_j \) for both spouses. The findings are presented in Figure \ref{fig:ATT gamma: EQ-ttl lazy}. As asserted in Proposition 4, the sign of the labor supply effect hinges on \( \gamma \).

Following this, I explore a series of robustness checks, comparing their outcomes with the baseline model in Figure \ref{fig:ATT gamma: EQ-ttl lazy}. It's important to understand that these checks are rooted in simulations using parameters from Table \ref{table: estimation results}; separate estimations for models within these checks haven't been conducted (further details in Appendix \ref{sec A: robusness checks}). It's also worth noting that the models in this section are notably more involved than the main one,  so while the robustness checks follow the same procedure as in the previous section, they incorporate several simplifications:

\begin{enumerate}
    \item All wives in the model are of the same age, equivalent to the sample's average age of \( 38.5 \).
    \item I only consider values of \( \gamma \) within the range \([0.8,1.2]\), as this zone is of primary interest and it minimizes the number of simulations.
    \item Since the models don't have a closed-form solution, all choices must be discretized.
    \item I simulate significantly fewer families.
\end{enumerate}

\textit{Note:} The plots in the robustness check models might appear less smooth due to the discretization and a reduced number of simulated families.
\begin{itemize}
\item \textbf{Endogenous divorce.} I allow the household to choose marital status in the second period. They get divorced only if they both consider divorce as a better option than marriage. The key change is that in the second period each spouse experiences a "love" shock and if it is negative enough for both spouses, they get divorced. I choose the distribution of love shocks so that the probability of divorce under TBR was close to $0.2$, the value I use in my main model. As it can be seen from Figure \ref{fig:ATT gamma: EQ-ttl-endog}, it has minimal effect on results shown on Figure \ref{fig:ATT gamma: EQ-ttl lazy}. And sign is still switched around the value of $\gamma$ equal to 1.

\item \textbf{Endogenous divorce and "bribing".} I further relax the main model by allowing "bribing". If, for instance, the wife wants to get divorced but the husband does not want to, she can try to "bribe" him. They look for a share which makes the husband indifferent and keeps the wife inclined to get divorced. If they find such a share, they stick to it and get divorced; if not, they stay married. If they both want to get divorced, they just use the share rule prescribed by the regime. Figure \ref{fig:ATT gamma: EQ-ttl-endog-bribe} shows that even this extension of the model does not change the main pattern from Figure \ref{fig:ATT gamma: EQ-ttl lazy}, having $\gamma=1$ as a pivotal point. 
\end{itemize}

The robustness checks indicate that extending the model not only has no qualitative effect on the positive correlation between the treatment effect size and risk aversion but also underscores the importance of the condition from Proposition 4. Both extensions reveal that the treatment effect of EDR on labor supply switches its sign from negative to positive at $\gamma=1$. This suggests that precautionary savings might be the sole reason for the positive effect of EDR on labor supply, even in settings with endogenous divorce and the ability of spouses to bargain with one another. However, it should be noted that the final conclusion cannot be derived solely from simulations. Therefore, subsequent versions of the paper will present precise conditions similar to those in Section \ref{sec:CRRA condition main}.

\subsection{Decomposition of Treatment Effect}

In this section, I will attempt to understand the contribution of two elements of EDR:
\begin{itemize}
    \item \textbf{Deviation from the optimal share:} Under TBR in the main model, the household assigns formal ownership in a way that divorce does not result in different consumption for any spouse, thereby completely eliminating the risk of divorce. Even if there is no court intervention under EDR, as long as the new share enforced by law differs from the one under TBR, it reintroduces uncertainty previously eliminated. This change necessitates additional action from the household in the form of increased savings to insure themselves against the risk of divorce.
    \item \textbf{Increased uncertainty due to court intervention:} Given that households cannot fully predict court decisions, court intervention increases overall uncertainty, pushing precautionary savings even further.
\end{itemize}

To understand which element of these two makes the main contribution to the treatment effect from Table \ref{table:ATT_5_years}, I simulate two counterfactual policies:
\begin{itemize}
    \item \textbf{C1: Deterministic Share Shift}. This policy assigns a fixed and commonly known share of savings to the wife, equal to the mean share under EDR, $E(\alpha)=0.49$.

    \item \textbf{C2: Variance Increase Only}. The policy assigns a random share while maintaining the mean at the TBR level. Essentially, it represents a mean-preserving transformation of TBR with increased variance from zero to $0.07$ (see Table \ref{table: estimation results}).
\end{itemize}

\begin{figure}
\caption{Share distribution under: EDR, C1, and C2}\label{fig: share policies.}
    \centering
    \includegraphics[scale=0.25] {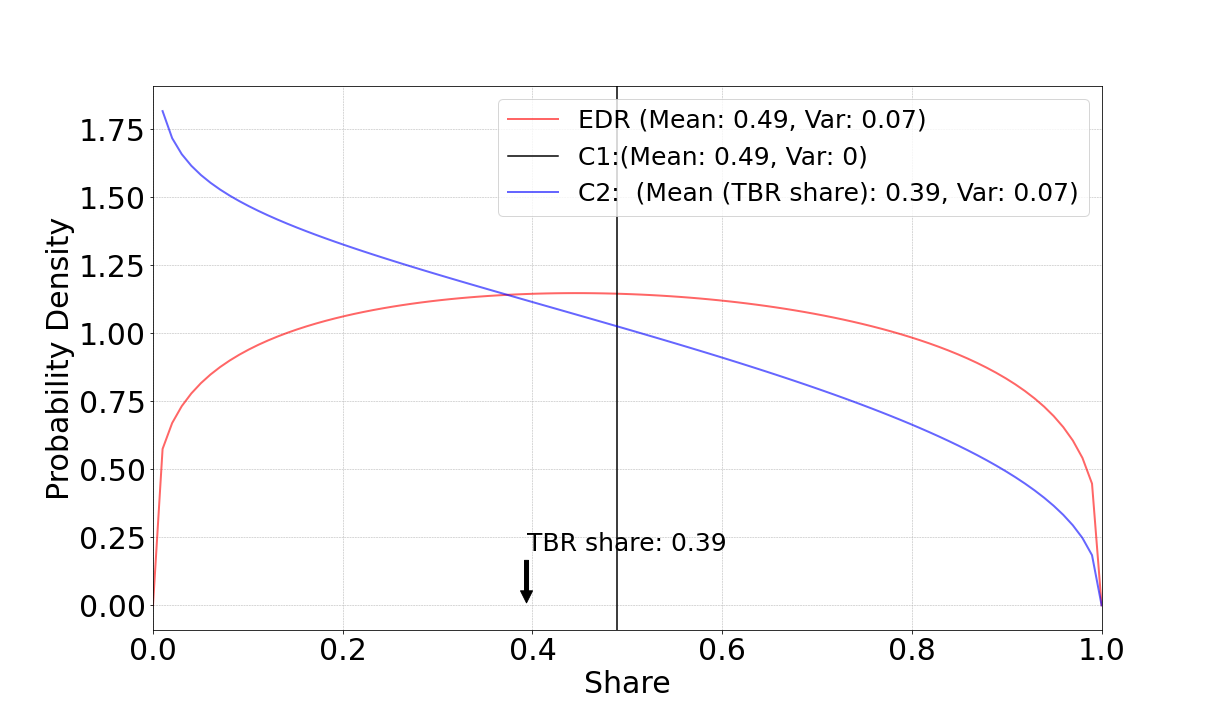}
    
    \label{fig:enter-label}
\end{figure}

Figure \ref{fig: share policies.} shows the distribution of the wife's share in the case of divorce under different policies. It can be seen that $C1$ is simply described by a fixed share, so I use a solid vertical line for it. While $C2$ is constructed in a way to maintain the same mean as under TBR, 0.41, it is not symmetrically centered around it and is skewed to the left. The distribution under EDR is almost symmetric around 0.5.

\begin{table}[]
    \centering
    \caption{Treatment decomposition (simulations).}\label{tab:Treatment decomposition}
    \begin{tabular}{c|ccc|}
    \toprule
     & \multicolumn{3}{c|}{\textbf{Transitions}} \\
     \midrule
    \textbf{Variable} &
    \multicolumn{1}{c|}{\bfseries \footnotesize $TBR \rightarrow C1$} &
    \multicolumn{1}{c|}{\bfseries \footnotesize $C1\rightarrow EDR$} &
    \multicolumn{1}{c|}{\bfseries \footnotesize $TBR\rightarrow EDR$} \\
    
    \midrule
         $l_f$ &            0.02 &          2.38 &         2.4 \\
     $l_m$ &            0.02 &          1.58 &         1.6 \\
$I[l_f>0]$ &          0.0008 &          0.05 &        0.05 \\
$I[l_m>0]$ &          0.0003 &          0.03 &        0.03 \\
   $s-S_0$ &             0.3 &            22 &        22.3 \\
    \midrule
    & \multicolumn{3}{c|}{\textbf{Transitions}} \\
    \midrule
     \textbf{Variable}  &
    \multicolumn{1}{c|}{\bfseries \footnotesize $TBR\rightarrow C2$} &
    \multicolumn{1}{c|}{\bfseries \footnotesize $C2\rightarrow EDR$} &
    \multicolumn{1}{c|}{\bfseries \footnotesize $TBR\rightarrow EDR$} \\
    \midrule
          $l_f$ &           7.8 &            -5.4 &         2.4 \\
     $l_m$ &           4.5 &            -2.9 &         1.6 \\
$I[l_f>0]$ &          0.16 &           -0.11 &        0.05 \\
$I[l_m>0]$ &          0.07 &           -0.04 &        0.03 \\
   $s-S_0$ &          69.7 &           -47.4 &        22.3 \\
   \bottomrule
    \end{tabular}
    \vspace{0.5em} 
    \centering
    \footnotesize{
    
    \textbf{Notes:} a) All numbers are simulation results.\\
    b) The first two transitions must sum up to the third one \\ by construction.
    }
\end{table}

It can be seen from Table \ref{tab:Treatment decomposition} that the main contribution to the positive effect of EDR on labor supply and savings is made by the randomness of the share rather than by deviation from the optimal level (see $(C1 \rightarrow EDR)$ and $(TBR \rightarrow C2)$). Moreover, the transition $(C2 \rightarrow EDR)$ produces negative results, which implies that increasing variance alone to the EDR level, while keeping the mean at the TBR optimal share level, $\alpha^*=0.41$, can result in a higher effect than the full transition $(TBR \rightarrow EDR)$. The negative effect of the transition $(C2 \rightarrow EDR)$ is explained by the shape of the underlying distribution for both policies and the utility function shape: Under $C2$, the wife often receives a share close to zero, as the distribution of $\alpha$ is skewed towards zero. Under the CRRA utility function, this leads to very low utility, since $u'(0)=\infty$. This means that $C2$ is associated with higher variance in utility, while the distribution of the share $\alpha$ under EDR does not produce extreme values of share for any of the spouses. This indicates that the transition $(C2 \rightarrow EDR)$ decreases uncertainty, allowing spouses to supply less labor and reduce precautionary savings. This decomposition shows that the main reason of people increasing their labor supply due to EDR was increase uncertainty caused by court intervention in sharing process.

\subsection{Welfare Effect.}

From the previously discussed results, the examination of how the policy tangibly influenced spouse welfare becomes ambiguous, as there are two competing welfare effects: 1) The initial goal of the policy is to increase the share of savings for the second earner (often the wife) in case of divorce, which should theoretically enhance the welfare of the spouse whose share is increased due to the policy. 2) However, EDR increases the uncertainty faced by the household, leading to a higher labor supply response. This increase in labor and corresponding decrease in leisure time during the marriage can offset the positive effect on the spouse who receives more savings in the event of divorce. For this reason, I use the estimated model to compute the welfare effect of EDR and the counterfactual policies $C1$ and $C2$

Welfare, \(W_j\), for each spouse is defined as:

\begin{equation}
    W_j = u(c_{1,j},l_j) + \beta E\Big[(1-\pi) U(c_{2,j})+\pi U(c_{2,j,d})\Big]
\end{equation}

A notable issue with the CRRA utility function is that it yields negative values for \(\gamma > 1\), complicating analyses of not only absolute but also relative welfare changes. For instance, if the utility of the wife is $-5$ under TBR and decreases to $-6$ under EDR, then the absolute change, $-1$, is not particularly informative. However, the analysis of relative change is also confusing, as it equal to \(\frac{-6 - (-5)}{-5} \times 100\% = 20\%\), which is greater than 0. This implies that even though the absolute change is negative, the relative change is positive. Consequently, I focus on the compensation, received as a lump-sum by the family at the onset of the first period, required to achieve the welfare level under TBR. Table \ref{tab: Welfare compensation} demonstrates the results of this analysis.

\begin{table}[h!]
    \centering
    \caption{Lump-Sum Compensation for Policy Transition from TBR}\label{tab: Welfare compensation}
    \label{tab:compensation}
    \begin{tabular}{lccc}
        \toprule
         &  EDR &  $C1$ & $C2$ \\
        \midrule
        Wife & \$26.12&         -\$1.75 &        \$231.73 \\
Husband & \$19.89 &          \$5.41 &         \$21.56 \\
 Family & \$21.22 &          \$3.26 &        \$106.61 \\
        \bottomrule
    \end{tabular}
    \footnotesize{
    
    \textbf{Notes:} "The table illustrates the financial cost to the government of equating each agent's utility, as shown in the left column, with the welfare under TBR. These figures can be interpreted as weekly lump-sum compensations, given the normalization of labor supply to weekly hours. The compensation is measured in 1990 US dollars. The average weekly labor income for wives in my sample is approximately \$71, and for husbands, it is about \$110}
\end{table}
From Table \ref{tab: Welfare compensation}, it is evident that despite the policy's intention to increase the wife's share in the event of a divorce, EDR actually decreases the expected welfare of the wife. This is because the government would need to provide the wife with positive compensation to maintain her welfare at the TBR level after the transition to EDR. This outcome arises more from the EDR's effect on the wife during the marriage than post-divorce. While EDR indeed increases the wife's consumption in the event of a divorce, it decreases her consumption and leisure during the marriage. Policymakers might have unintentionally overlooked such household behavior as precautionary savings, which subsequently reduced the wife's welfare.

An important observation emerges from the implementation of counterfactual policy $C1$. This policy allocates a fixed share of \$0.49 to the wife, contrasting with the \$0.39 share she receives under TBR. Notably, this adjustment results in a less pronounced decline in the husband's welfare.\footnote{The $C1$ policy mirrors the Community Property regime, as adopted in states like California and Texas, where household property is evenly divided between spouses.} It should be noted, however, that this policy is unlikely to positively affect the husband's welfare. This is due to the dual negative impacts they face: increased uncertainty during the marriage as part of the household dynamic, and a reduced share of savings in the event of a divorce. In contrast, $C1$ has the potential to enhance the wife's welfare. The policy induces only a marginal increase in uncertainty, while ensuring that the wife's average share remains consistent with that under EDR

On the other hand, the counterfactual policy $C2$ results in a significant welfare decline for the wife. As explained previously, this can be attributed to the shape of the share distribution under $C2$, which is heavily skewed toward zero, as depicted in Figure \ref{fig: share policies.}. Consequently, the wife expects her consumption to be close to zero with a high probability in the event of a divorce. Given the properties of the CRRA-utility function, which equates to negative infinity when the consumption value is zero and $\gamma > 0$, this situation leads to a substantially decreased expected utility.

It's important to emphasize that the adverse impact of EDR on household welfare is not surprising, due to the policy's constraining nature. Instead of receiving a share that would maximize household welfare, spouses are allocated a different share that may not necessarily be the optimal choice. Furthermore, even if $C2$ offers a mean share equivalent to the optimal share under TBR, it's evident that as long as there's a positive probability of receiving a non-optimal share, the welfare effect must be negative.

\section{Concluding Remarks.}\label{sec: conclusion}
This research, employing recent staggered DiD methodologies, investigates the effects of the Equitable Distribution Regime (EDR) on household behavior. The first key finding is that EDR, designed to increase the property share of wives in the event of a divorce, led to an increase in female labor supply. The second key insight is the underlying mechanism driving this effect: a) Under the earlier Title-Based Property Regime (TBR), households could choose and commit to each spouse's property share in the case of a potential future divorce, simply by assigning property ownership. This allowed households to completely smooth consumption across two possible future marital statuses: remaining married or getting divorced. b) The introduction of EDR removed this choice from households. Without the option to secure their consumption in advance against potential divorce shocks, households faced increased risks. They responded as prudent agents, with higher savings, which was financed by increased labor supply.

I underscore the significance of this precautionary savings motive by demonstrating that the effect of EDR on labor and savings is determined solely by prudence. Introducing extensions, such as endogenous divorce or the ability to bargain over the share of assets for each spouse in the event of a divorce, does not alter this conclusion.

Further analysis exposes a perverse effect of the policy. While EDR increased the post-divorce welfare of wives, it might have inadvertently decreased their welfare during marriage. Crucially, court intervention in property allocation is the primary channel through which EDR increased labor supply and decreased the welfare of both spouses. Thus, for policymakers looking to enhance the wife's share of assets post-divorce, adopting a policy similar to the Community Property Regime — where property is equally divided — would be more effective. The Community Property Regime minimally impacts household welfare while successfully increasing the wife's share of assets.

\clearpage

\printbibliography

\clearpage

\begin{appendix}
\setcounter{table}{0}
\setcounter{figure}{0}

\renewcommand{\thetable}{A.\arabic{table}}
\renewcommand{\thefigure}{A.\arabic{figure}}

\section{Marital Assets Distribution Under TBR and EDR: New York Case}\label{sec: Asset distibution NY}

\begin{table}[h!]
\centering
\caption{Distribution of Marital Assets in the Contested Case Sample (Couples with Marital Assets), by Year and Case Group}\label{tab: asset distribution law }
\begin{tabular}{@{}lcccc@{}}
\toprule
\textbf{Cases Group} & 1978 & 1984 & \textbf{Difference (p.p.)} \\
 & \% & \% &  \\
\toprule
\textbf{Cases with Complete Asset Information} & \textbf{(n=84)} & \textbf{(n=68)} &  \\

Majority to Wife        & 41 & 49 & +8  \\
Relatively Equal        & 11 & 19 & +8  \\
Majority to Husband     & 49 & 32 & -17 \\
Median \% to Wife       & 45 & 69 & +24 \\
Average \% to Wife      & 49 & 58 & +9  \\
Wife Received 50\% or More & 49 & 60 & +11 \\
\addlinespace
\textbf{Valid Group}    & \textbf{(n=140)} & \textbf{(n=130)} & \\
\midrule
Majority to Wife        & 39 & 40 & +1  \\
Relatively Equal        & 16 & 26 & +10 \\
Majority to Husband     & 45 & 34 & -11 \\
Median \% to Wife       & 50 & 51 & +1  \\
Average \% to Wife      & 49 & 51 & +2  \\
Wife Received 50\% or More & 51 & 52 & +1  \\
\bottomrule
\end{tabular}

\textit{Note:}
\footnotesize a) Majority = More than 60\% b) Relatively equal = Between 40\% and 60\% c) New York adopted Equitable Distribution Regime (EDR) in 1981.  d) Data taken from \textcite{garrison1991good}.
\end{table}

\textit{Note:} The data from this section was not available during the  model estimation stage and, as such, was not included list of moments. It will, however, be considered for incorporation in future iterations of the paper.

The \textcite{garrison1991good} defines the \textit{Valid Group} as a subset of the contested case sample which has been used for the analysis of property and debt division within their study. This subset adheres to several criteria:

\begin{itemize}
    \item The \textit{Valid Group} includes cases that either have complete information on assets and debts (\textit{A Group}) or have comprehensive distributional data for at least 90\% of the known pool of assets and/or debts with no more than one missing valuation (\textit{B Group}).
    \item Cases with substantial assets lacking reliable valuations, such as the marital home, other real estate, businesses, or pensions, were excluded from the \textit{Valid Group} due to their significant impact on the financial analysis.
    \item To prevent distortion in the distribution percentages, cases with negative net worth were omitted from most of the analyses, as these could convey misleading implications about the financial outcomes for the parties involved.
    \item The \textit{Valid Group} was further scrutinized against the full contested sample to ensure its representativeness. This involved making specific assumptions for missing data and checking the consistency of results.
    \item A detailed set of analysis protocols was employed for cases in the \textit{B Group}. This protocol involved the exclusion of any assets or debts with missing valuations from the analysis and the equal distribution of assets or debts with missing distributions between spouses.
\end{itemize}

The author's selection criteria for the \textit{Valid Group} allow for a more precise and representative analysis of the impact of New York's Equitable Distribution Law on marital property  distribution.

\begin{table}[h!]
\centering
\caption{Proportion of Cases When Wives Receive 50\% or More by Asset Type}
\label{tab: asset_type_distribution_law}
\begin{tabular}{@{}lccc@{}}
\toprule
\textbf{Asset Type} & \textbf{1978} & \textbf{1984} & \textbf{Difference (p.p.)} \\
 & \% (N)& \% (N) & \\

\midrule
Automobile & 44 (507) & 47 (599) & +3 \\
Business Interests & 12 (50) & 20 (70) & +8\\
Household Goods & 67 (240) & 70 (209) & +3 \\
Jewelry & 77 (71) & 77 (97) & 0 \\
\addlinespace
\textbf{Liquid Assets:} & & & \\
\hspace{3mm}a) Bank Accounts & 54 (686) & 56 (722) & +2 \\
\hspace{3mm}b) Other & 42 (273) & 48 (267) & +6 \\
\addlinespace
\textbf{Nonliquid Assets:} & & & \\
\hspace{3mm}a) Pensions & 15 (41) & 31 (85) & +16 \\
\hspace{3mm}b) Other & 26 (99) & 56 (159) & +30 \\
\addlinespace
Marital Residence & 75 (288) & 72 (304) & -3 \\
Other Real Estate & 37 (116) & 40 (160) & +3 \\
Other Assets & 35 (207) & 45 (231) & +10 \\
\bottomrule
\end{tabular}

\footnotesize\textit{Note:} a) New York adopted Equitable Distribution Regime (EDR) in 1981. \\b) Data taken from \textcite{garrison1991good}.
\end{table}

Table \ref{tab: asset_type_distribution_law} illustrates the impact of the EDR on the proportion of cases where wives received 50\% or more of assets by asset type. The adoption of EDR appears to have generally increased the share for wives across various asset types. The only asset category that saw a reduction in the share of wives receiving at least half was the Marital Residence.

\section{Staggered Adoption DiD.}\label{sec: Diff-in-Diff description}
Difference-in-Differences (DiD) methods are fundamental in estimating causal effects within observational studies, facilitating the comparison of outcome shifts over time between treated and control groups. \textcite{borusyak2021revisiting} and \textcite{callaway2021difference} introduced an approach to estimate causal effect in staggered adoption setting. In this section I provide high level description of both  methods
\subsection{Main method: Borusyak et al, 2021 (BJS)}
Methods estimates the causal effects of a binary treatment $D_{it}$ on an outcome $Y_{it}$, considering a staggered adoption design in a panel of units $i$ and periods $t$. Here, treated status is an absorbing state, where once a unit is treated, it remains treated. The method is based on two main assumptions:
\subsubsection{Assumptions.}
\begin{enumerate}
    \item \textbf{Parallel trends:} There exist non-stochastic $\alpha_i$ and $\beta_t$ such that 
\[ E [Y_{it} (0)] = \alpha_i + \beta_t \]
for all $it \in \Omega$.
\item \textbf{No anticipation effects:}
\[ Y_{it} = Y_{it}(0) \]
for all $it \in \Omega_0$.
\end{enumerate}

\subsubsection{Estimation Algorithm}
First method estimates ``individual" treatment, $\tau$
\begin{enumerate}
     \item Combine all untreated ("not-yet treated") units and estimate:
    \begin{equation}
        Y = A' \lambda + X' \delta + \varepsilon.
    \end{equation}
    where $X$ includes regular covariates and $A$ consists of fixed effects
    \item For every treated unit, compute $\hat{\tau}_{i,t}=Y_{it} -\hat{Y}_{i,t}(0)$, where $\hat{Y}_{i,t}(0)=A' \hat{\lambda} + X' \hat{\delta}$.
    \item Apply $\hat{\tau}_{i,g+r}$ for different aggregations, as illustrated by BJS, to get an aggregate average treatment effect (ATE) for various periods post-intervention.
\end{enumerate}

\subsection{Secondary method: Callway and Sant'Anna, 2021}

\subsubsection{ Assumptions}
To effectively employ the DiD method, the following foundational assumptions are indispensable:
\begin{enumerate}
    \item \textbf{Conditional Parallel Trends Assumption:} In the absence of the treatment, both treated and control groups would follow parallel trajectories, conditional on observed covariates.
    \item \textbf{SUTVA (Stable Unit Treatment Value Assumption):} The treatment status of a unit does not affect the outcomes of other units.
    \item \textbf{No Anticipation:} Units do not alter their behavior in anticipation of future treatments.
    \item \textbf{Overlap:} The probability of being treated must be strictly between $0$ and $1$.
    \item \textbf{Irreversibility of Treatment:} Once a unit is treated, it always remains treated.
    \item \textbf{Random Sampling:} Broadly, this implies that all variables $(y_i,x_i,D_i)$ are independently and identically distributed across families. For instance, if $x_i$ represents years of schooling, it means that educational years for every period are drawn simultaneously.
\end{enumerate}

\subsubsection{Estimation Algorithm}
The method systematically accommodates variations in treatment timing via the subsequent steps:
\begin{enumerate}
    \item \textbf{Group Formation:} Divide the data into unique groups based on their respective initiation times of treatment.
    \item \textbf{Average Treatment Effects Estimation:} Determine the average treatment effects for each group and time period, often using regression adjustment to account for covariates.
    \begin{itemize}
        \item Select a group treated in year $g$ and determine the time length $k$ after the treatment for which you want to estimate the treatment effect. For example, to estimate the effect for the first 5 years after adopting EDR.
        \item Choose all observations treated after time period $g+k$, using them as the control group for group $g$.
        \item Estimate the OLS regression difference of the outcome of interest, $Y_{g+\tau}-Y_{g-1}$, using the control group. Use the estimated function $\hat{m}$ to impute potential outcomes for treated observations.
        \item Estimate the generalized propensity score, $p_{g,t}(X)$: the probability of being treated in period $g$\footnote{Generalized propensity score is the probability of belonging to some particular group, based on year of adoption EDR.}.
        \item Construct the estimator. In this paper, a Doubly-Robust estimator is utilized. Formally,
        \begin{equation}
            ATT(g,t)=E[W(Y_t-Y_{g-1}-\hat{m}_{g,t}(X)) ]
        \end{equation}
        \begin{equation}
            W=\left(\frac{G_g}{E(G_g)}-\frac{\frac{p_{g,t}(X)(1-D_t)(1-G_g)}{1-p_{g,t}(X)}}{E\left[\frac{p_{g,t}(X)(1-D_t)(1-G_g)}{1-p_{g,t}(X)}\right]}\right)
        \end{equation}
    \end{itemize}
    \item \textbf{Weighted Averaging:} Compute a weighted average of the group-specific treatment effects, $ATT(g,t)$, to infer an overall average treatment effect.
    \item \textbf{Inference:} Execute statistical inference on the estimated effects using methods like bootstrap techniques.
\end{enumerate}

\clearpage
\section{Verification of Assumptions  for DiD.}\label{sec: Assumption test}
\begin{figure}[h!]
    \centering
    \caption{Propensity Score Distributions for Different Adoption Groups}\label{fig: Propensity score}
    \label{fig:propensity}
    
    \begin{subfigure}{0.3\textwidth}
        \caption{) Group 1974}
        \includegraphics[width=\linewidth]{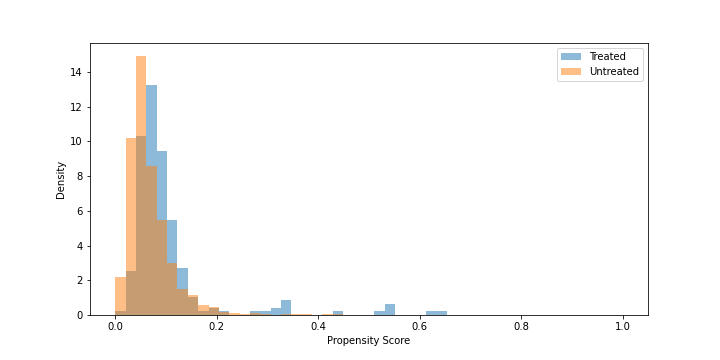}
    \end{subfigure}\hfill
    \begin{subfigure}{0.3\textwidth}
        \caption{) Group 1977}
        \includegraphics[width=\linewidth]{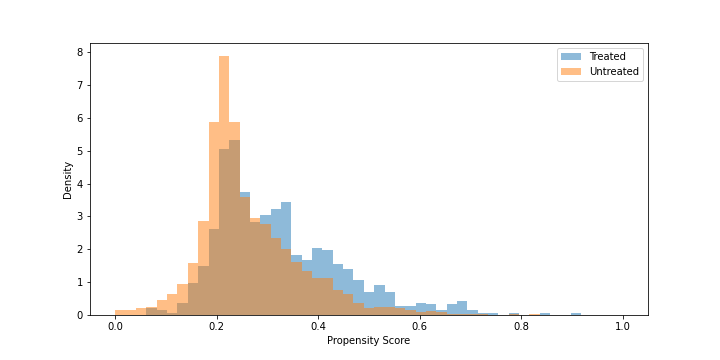}
    \end{subfigure}\hfill
    \begin{subfigure}{0.3\textwidth}
        \caption{) Group 1978}
        \includegraphics[width=\linewidth]{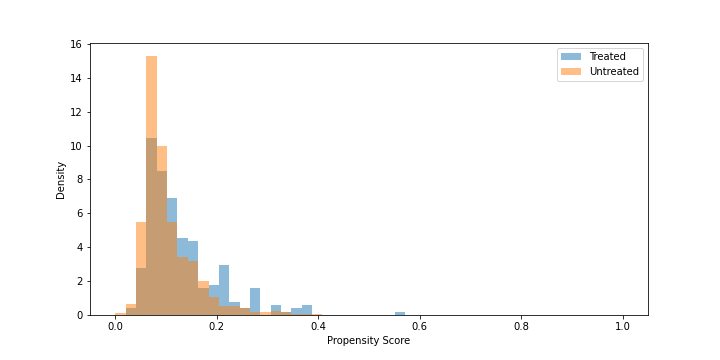}
    \end{subfigure}
    
    \medskip
    \begin{subfigure}{0.3\textwidth}
        \caption{) Group 1980}
        \includegraphics[width=\linewidth]{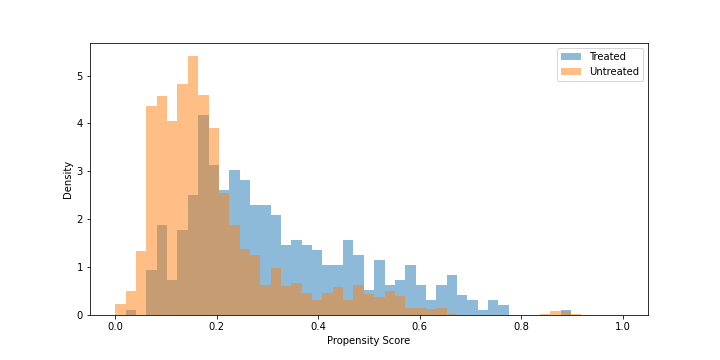}
    \end{subfigure}\hfill
    \begin{subfigure}{0.3\textwidth}
        \caption{) Group 1981}
        \includegraphics[width=\linewidth]{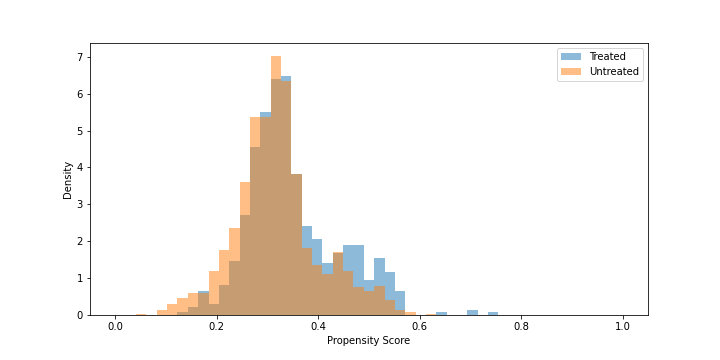}
    \end{subfigure}\hfill
    \begin{subfigure}{0.3\textwidth}
        \caption{) Group 1982}
        \includegraphics[width=\linewidth]{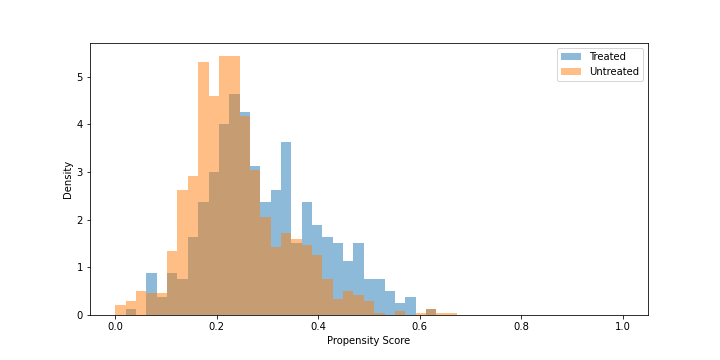}
    \end{subfigure}
    
    \footnotesize\textit{Note:} Propensity score estimated using logistic regression.
\end{figure}

While assumptions are inherently unverifiable, it is possible to argue in favor or against some of them, as elucidated below:
\begin{itemize}
    \item \textbf{Conditional Parallel Trends Assumption:} The placebo tests of pre-policy periods in Figures \ref{fig:wife_labour_supply_effect} and \ref{fig:husband_labor_supply_effect} indicate that the conditional parallel trend assumption could not be rejected in the pre-policy period. Because the coefficients for pre-policy periods are all insignificant.
    \item \textbf{SUTVA:} This assumption is challenging to test. However, one mechanism through which EDR can affect control states—people moving between states in response to policy—is mitigated by focusing on non-movers, thereby ensuring that policy contributions are minimal. 
    \item \textbf{No Anticipation:} While this study leverages a stronger assumption than the original paper, parallel trend arguments showcase that households do not alter their behavior in anticipation of EDR, given the statistical indistinguishability of control and treated groups prior the policy adoption. 
    \item \textbf{Overlap:} The validity of this assumption is  evaluated by estimating the propensity score across various groups, differentiated by the timing of EDR adoption. Although the scores are determined for each year within a 5-year period post-EDR adoption, they are combined into a singular plot, as depicted in Figure \ref{fig: Propensity score}. It can be seen that the propensity score is concetrated away from 0 and 1. Even with the policy being adopted at the state level, the model effectively discriminates treated units from controls, as treated units consistently exhibit a higher propensity score. In general, the estimated propensity score ensures a robust overlap, allowing for robust and unbiased estimation \footnote{It’s crucial to underline that, while the pronounced overlap is present, it might stem from the model’s limited ability to  differentiate between treated and control units.}.

    \item \textbf{Irreversibility of Treatment:} This assumption holds at the state level as states continue to adopt EDR. At the individual level, it may not hold due to potential relocations; thus, these instances are not considered.
    \item \textbf{Random Sampling:} This hard-to-test assumption permits authors to consider potential outcomes and can be regarded as a relatively weak assumption. 
\end{itemize}

\begin{figure}
    \centering
    \caption{Event Study: Savings}\label{fig: Event Study savings}
    \includegraphics [scale=0.3]{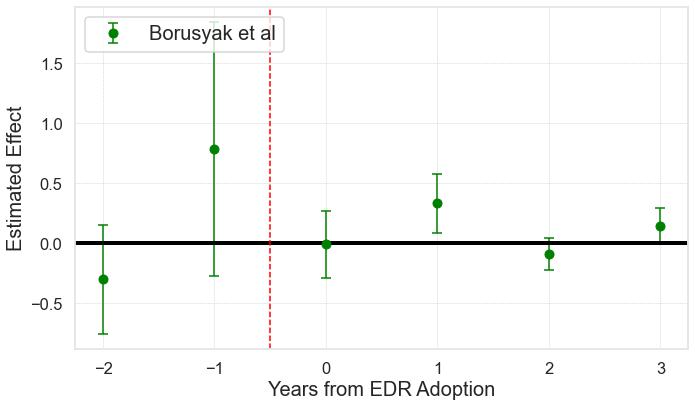}
    \label{fig:enter-label-2}
    \caption*{\footnotesize \textit{Note}:  a) FE included  marriage duration, husband's age, and wife's age, and  education beyond high school. b) Confidence intervals have a 90\% coverage.}
\end{figure}

\begin{figure}[h!]
\centering
\caption{Conditional Parallel Trend Assumption Test (Wife)}\label{fig:CPT wife_labour }
\begin{subfigure}{.5\textwidth}
\centering
\includegraphics[width=1\linewidth]{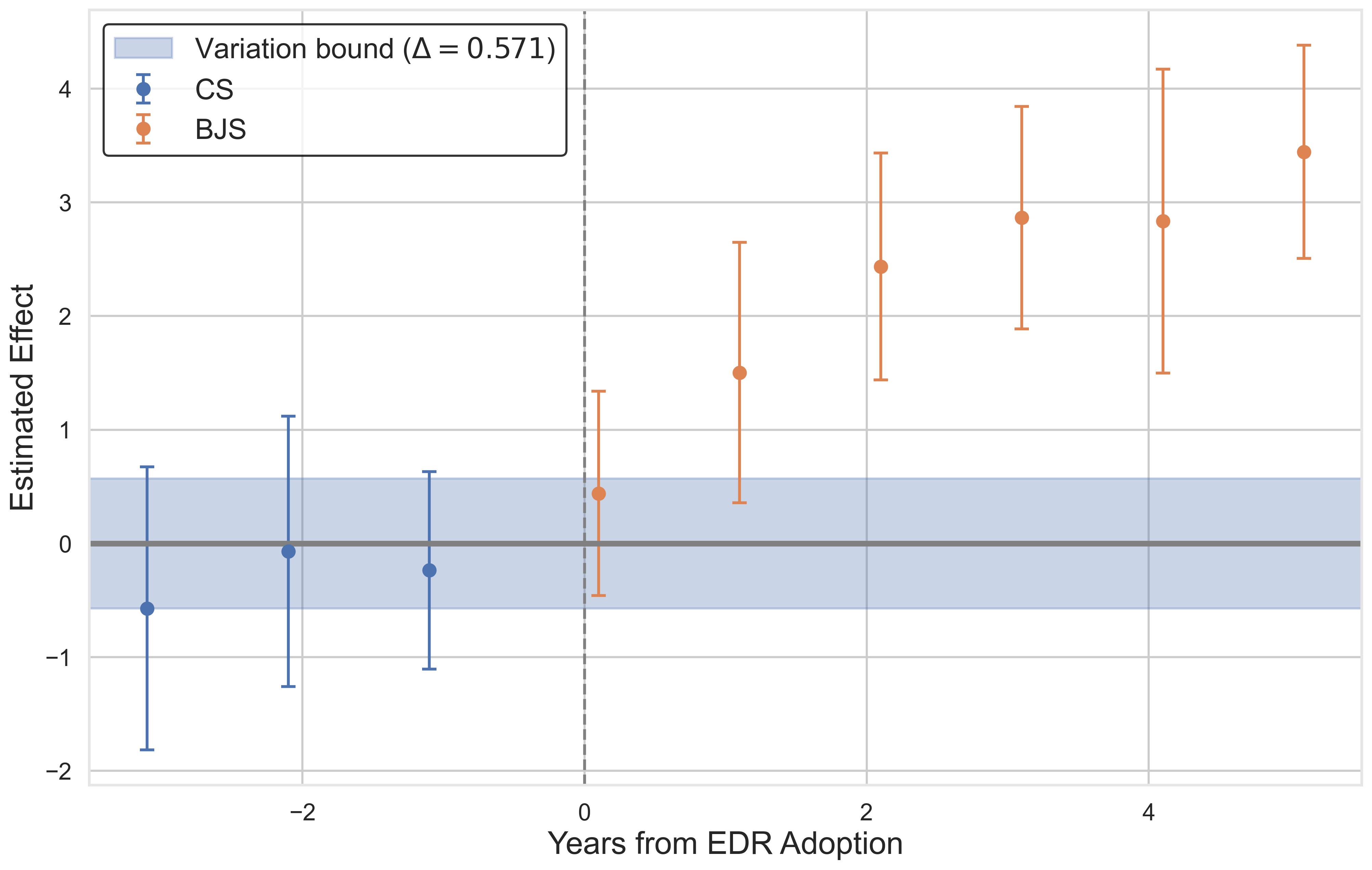}
\caption{) Weekly work hours}
\label{fig:wf_work_hours-2}
\end{subfigure}%
\begin{subfigure}{.5\textwidth}
\centering
\includegraphics[width=1\linewidth]{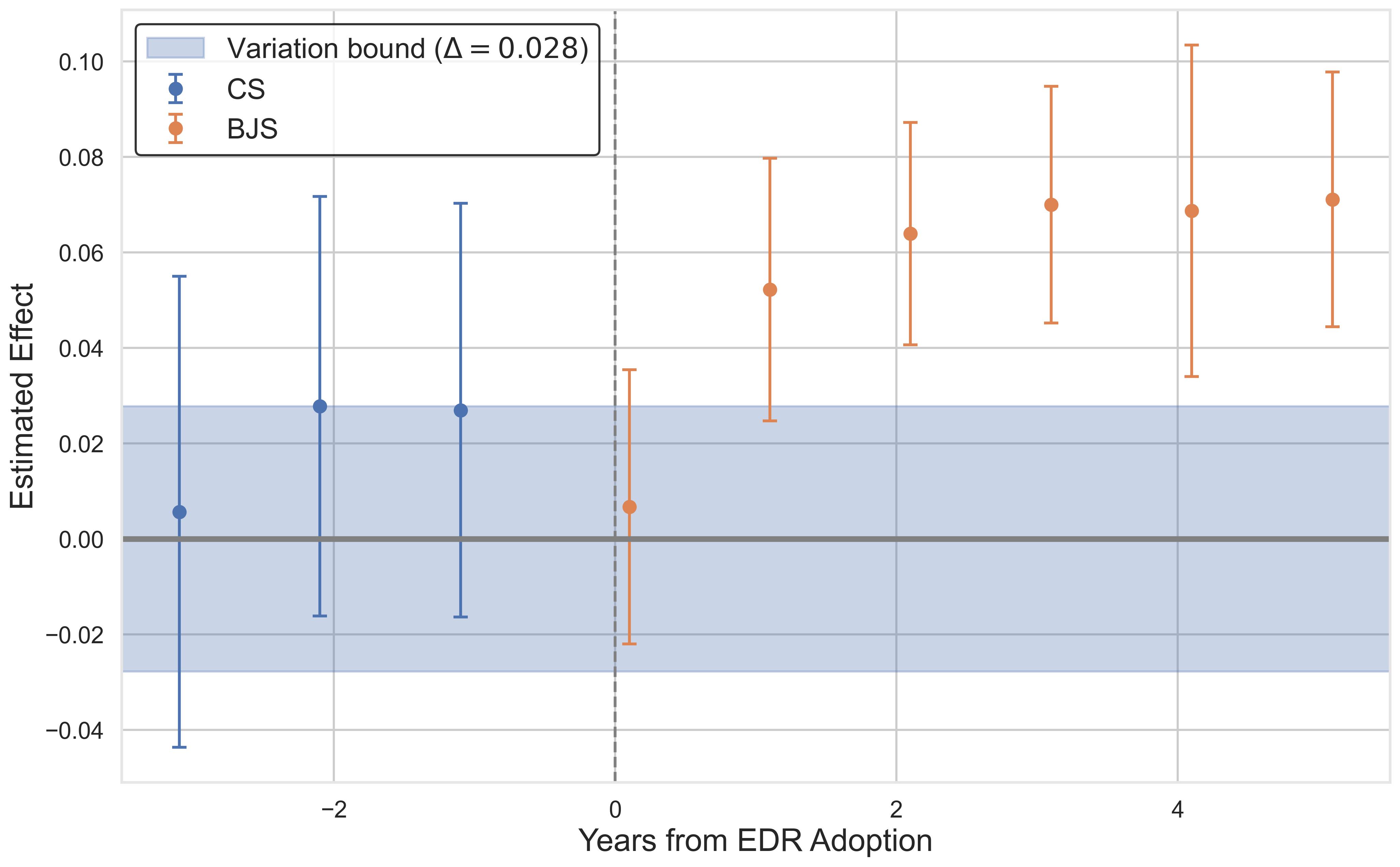}
\caption{) Employment (dummy)}
\label{fig:wf_work_dummy-2}
\end{subfigure}
\caption*{\footnotesize \textit{Note}: \footnotesize   a)  Blue coefficients are produced by CS methods, it includes following covariates: 2nd order polynomials of : age of spouses, duration of marriage, number of children; and indicator of having education higher then high-school for both spouses. O  b) Confidence intervals have a 95\% coverage. c) Families observed for fewer than 5 years post-treatment are excluded from the treatment group to avoid attrition bias.\footnote{They can, however, serve as control groups.} d) \textbf{Variation Bound} uses the higher in absolute value coefficient from pre-treatment period as $\Delta$}

\end{figure}

\begin{figure}[h!]
\centering
\caption{Conditional Parallel Trend Assumption Test (Husband)}\label{fig:CPT husband_labour }
\begin{subfigure}{.5\textwidth}
\centering
\includegraphics[width=1\linewidth]{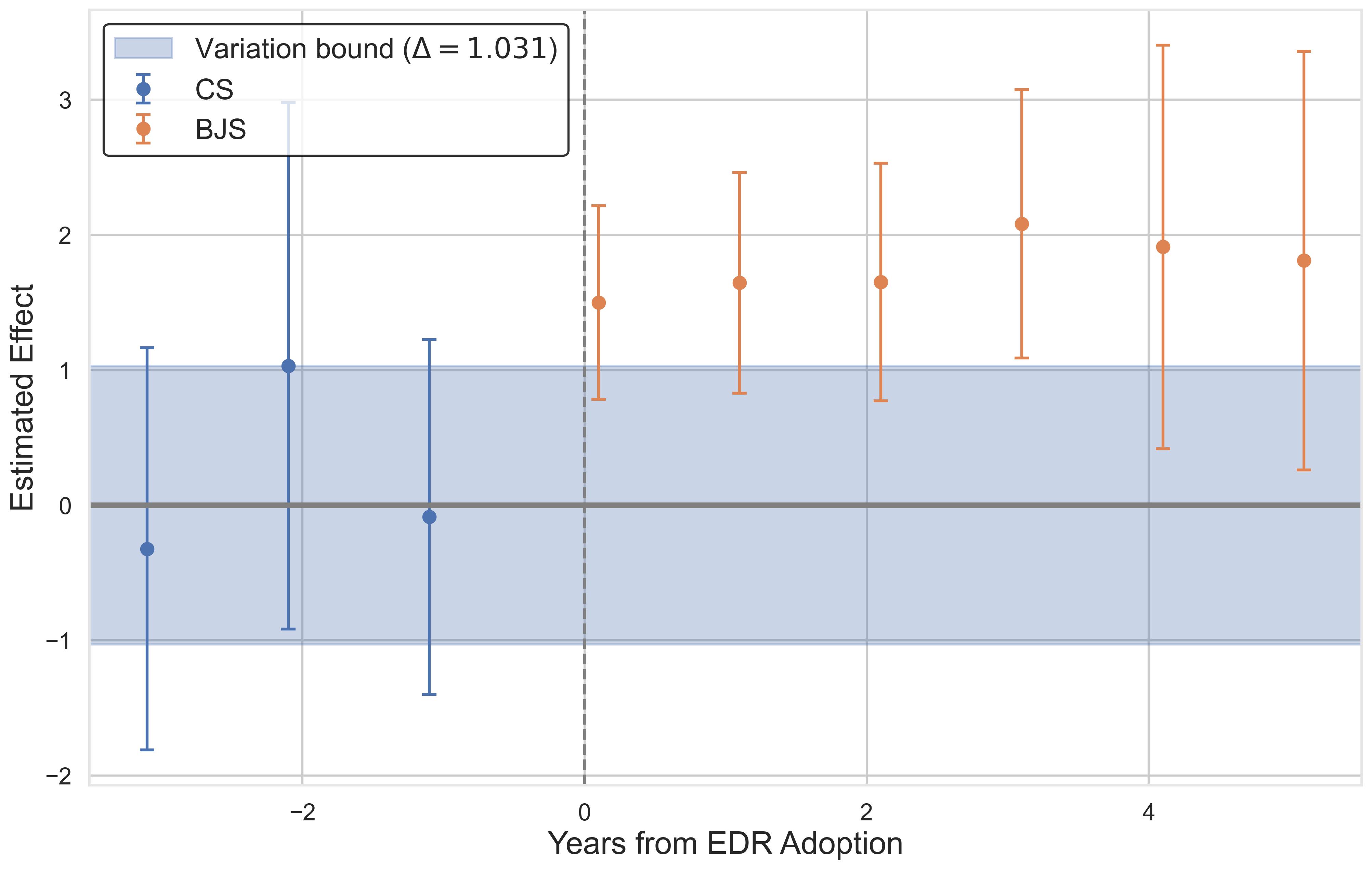}
\caption{) Weekly work hours}
\label{fig:wf_work_hours-3}
\end{subfigure}%
\begin{subfigure}{.5\textwidth}
\centering
\includegraphics[width=1\linewidth]{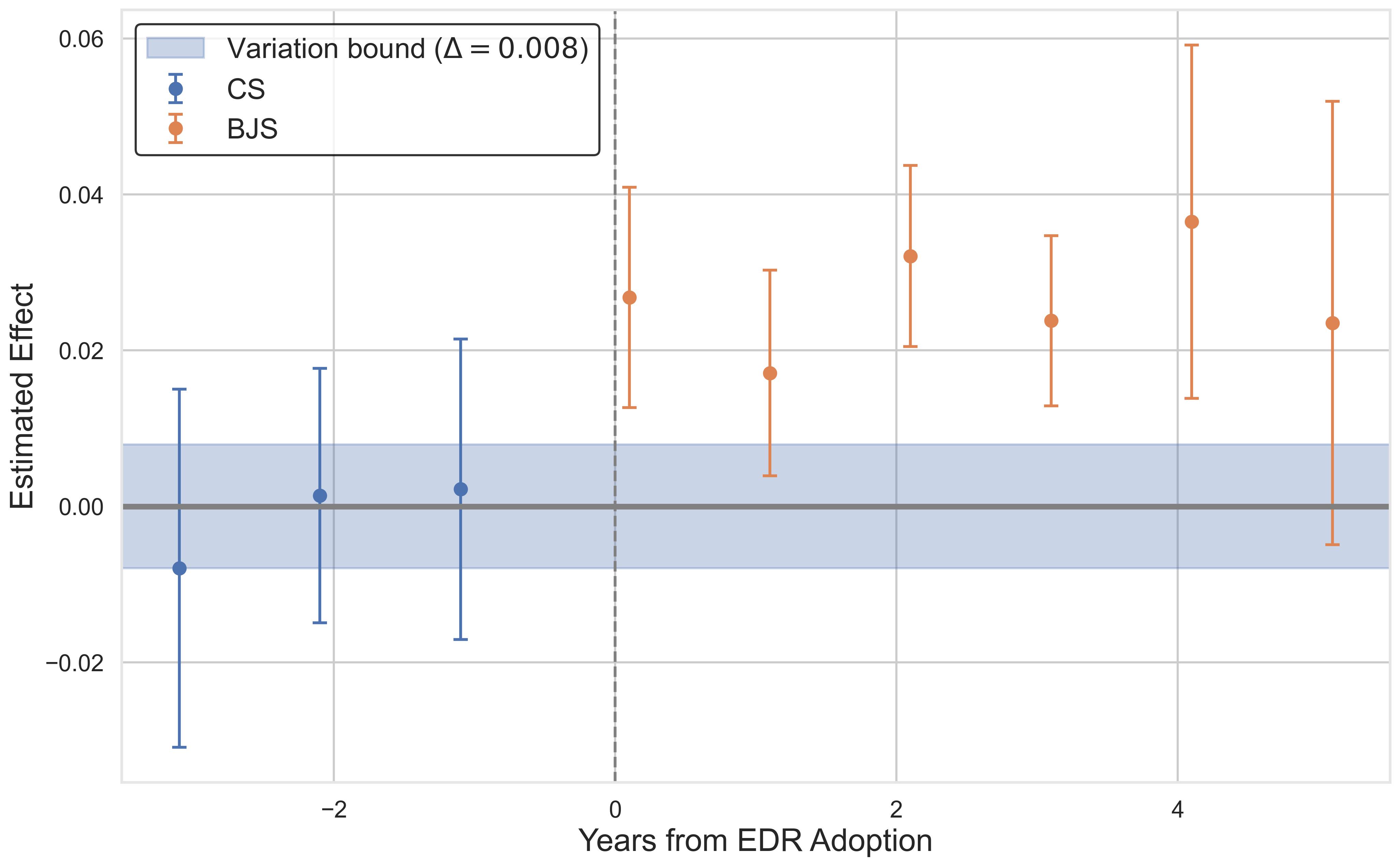}
\caption{) Employment (dummy)}
\label{fig:wf_work_dummy-3}
\end{subfigure}
\caption*{\footnotesize \textit{Note}: \footnotesize   a)  Blue coefficients are produced by CS methods, it includes following covariates: 2nd order polynomials of : age of spouses, duration of marriage, number of children; and indicator of having education higher then high-school for both spouses. O  b) Confidence intervals have a 95\% coverage. c) Families observed for fewer than 5 years post-treatment are excluded from the treatment group to avoid attrition bias.\footnote{They can, however, serve as control groups.} d) \textbf{Variation Bound} uses the higher in absolute value coefficient from pre-treatment period as $\Delta$}

\end{figure}

\clearpage
\section{Effect of EDR on labor supply (both methods).}

\begin{figure}[h!]
\centering
\caption{Effect of EDR on labor supply of wife (both methods)}\label{fig: both methods wife}
\begin{subfigure}{.5\textwidth}
\centering
\includegraphics[width=1\linewidth]{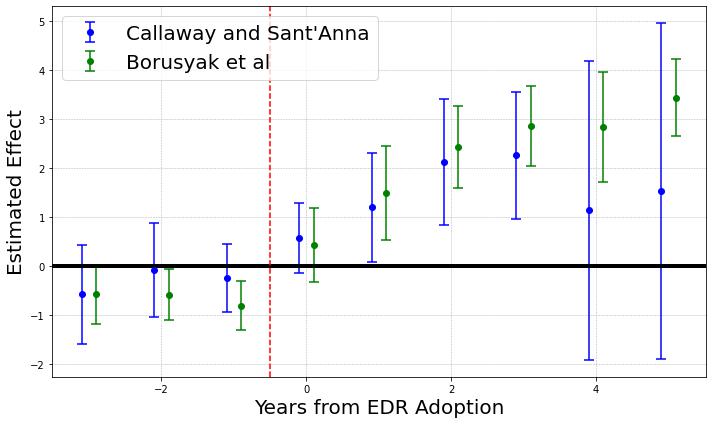}
\caption{) Weekly work hours}
\label{fig:wf_work_hours-4}
\end{subfigure}%
\begin{subfigure}{.5\textwidth}
\centering
\includegraphics[width=1\linewidth]{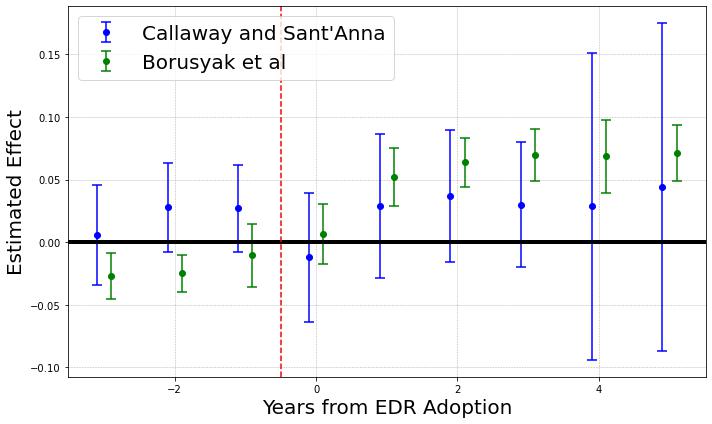}
\caption{) Employment (dummy)}
\label{fig:wf_work_dummy-4}
\end{subfigure}
\caption*{\footnotesize \textit{Note}: \footnotesize   a) Controls :  2nd order polynomials of marriage duration, husband's age, wife's age,  N. of children and an indicator for education    beyond high school. In case of BJS same variables are used as fixed effects. b) Confidence intervals have a 90\% coverage. c) Families observed for fewer than 5 years post-treatment are excluded from the treatment group to avoid attrition bias.\footnote{They can, however, serve as control groups.}}
\end{figure}

\begin{figure}[h!]
\centering
\caption{Effect of EDR on labor supply (both methods).}\label{fig: both methods husband}
\begin{subfigure}{.5\textwidth}
\centering
\includegraphics[width=1\linewidth]{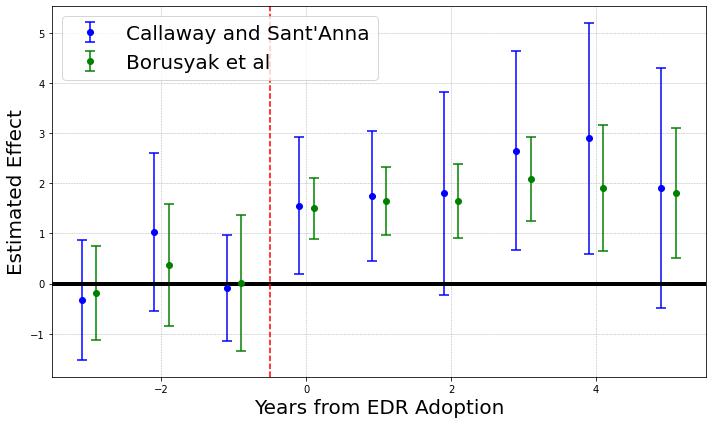}
\caption{) Weekly work hours}
\label{fig:hd_work_hours-2}
\end{subfigure}%
\begin{subfigure}{.5\textwidth}
\centering
\includegraphics[width=1\linewidth]{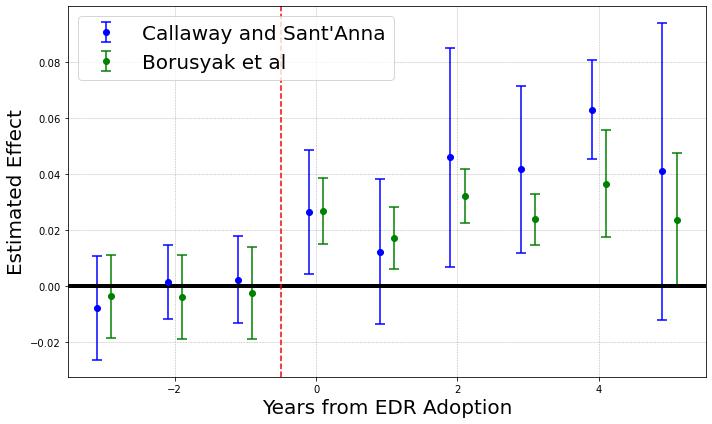}
\caption{) Employment (dummy)}
\label{fig:hd_work_dummy-2}
\end{subfigure}
\caption*{\footnotesize \textit{Note}:  a) Controls :  2nd order polynomials of marriage duration, husband's age, wife's age,  N. of children and an indicator for education    beyond high school. In case of BJS same variables are used as fixed effects. b) Confidence intervals have a 90\% coverage.c) Families observed for fewer than 5 years post-treatment are excluded from the treatment group to avoid attrition bias.\footnote{They can, however, serve as control groups.}}

\end{figure}

\clearpage
\section{Effect of EDR on Divorce Rate}\label{section: A Effect of EDR on Divorce Rate}
A core assumption driving our model is the exogeneity of divorce. This assumption might be overly simplistic if divorce rates are affected by policy changes. If such an influence exists, it could introduce another channel through which our observed results manifest. To scrutinize this potential effect, I have deployed a Difference-in-Differences (DiD) framework, mirroring the primary approach of the main body of the paper.

My original sample, being exclusive to married women, is not applicable for this specific investigation. Consequently, I have compiled a dataset encompassing all women, regardless of their marital status. This dataset incorporates an indicator variable that reveals whether a woman has ever divorced during her period of observation.

To ensure consistency with the primary focus of the paper, I have narrowed the sample to women who married before the initiation of the EDR policy. This deliberate constraint is designed to avoid complexities that might arise if the policy impacted marriage formations—since unions established post-policy might inherently differ from those formed prior.

However, here's where the nuanced difference in this analysis lies: To be able to test the parallel trend assumption using pre-treatment data, I include both women who remained married until the policy’s adoption and those who married but underwent a divorce before the policy. If the sample only contained women who were continuously married up until the policy’s introduction, the outcome variable would remain constant for all observations before the policy's adoption. This would make the parallel trend assumption test for the pre-treatment period trivial, as it would hold mechanically.

\begin{figure}[h!]
    \centering
    \caption{Effect of EDR on the Probability of Divorce.}\label{fig:female_divorce}\includegraphics[width=0.7\linewidth]{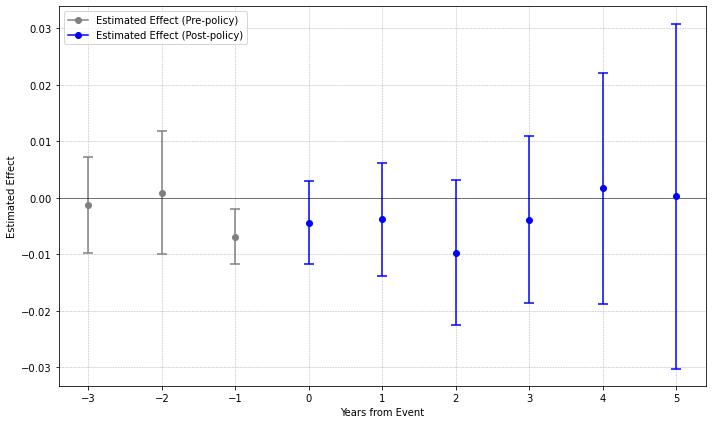}
    \caption*{\footnotesize \footnotesize \textit{Note:} a) Controls include a second polynomial of age and years of schooling. b) Sample consists of women aged 22 to 65 and includes states which never adopted Unilateral divorce while excluding states with community property regimes. c) Errors are clustered at the state level. d) Confidence intervals represent 90\% coverage.}
    
\end{figure}

Figure \ref{fig:female_divorce} showcases that the EDR policy appears to have a minimal impact on divorce rates. This observation supports the primary assumption that changes in labor supply are not the consequence of shifts in divorce probabilities.

To reiterate, the alignment between this sub-study and the main paper’s methodology is the focus on women who were wedded before the EDR’s enactment. By incorporating both women who remained married and those who divorced before the EDR’s adoption, I ensure sufficient variability in the outcome, allowing for a meaningful assessment of the parallel trends assumption during pre-treatment periods.

\clearpage
\section{Precautionary Savings and EDR.}

In this section I derive  conditions for EDR to increase savings. \textcite{lusardi1997precautionary} shows that wealth shock increases (stay the same or goes down) the savings of single agent if and only if relative prudence , $RP>2 (RP\leq 2)$ where $RP=-s\frac{u'''(x)}{u''(x)}$. To my knowledge there is no such derivation for my case with household instead of single agent. So I present the sufficient condition for increase (decrease of savings) in savings due to EDR.  
\subsection{Sufficient Condition for EDR to Increase (Decrease) Savings.}\label{sec: Sufficient condition}

The main difference between the TBR and EDR regimes in the household decision-making model is that under TBR, households can endogenously choose the share of each spouse in case of divorce, meaning that conditional on marital status in the second period, spouses have no uncertainty. The section is structured as follows:

\begin{itemize}
    \item Derivation of the First Order Condition (FOC) for the household problem under TBR.
    
    \item Investigation of the condition under which any deviation from the optimal share under TBR, $\alpha^*$, will lead to higher savings.
\end{itemize}

Let define the household problem.
\begin{align*}
\max_{c_{1,f}, c_{1,m}, c_{2,f},c_{2,m}, s, l_{1,f}, l_{1,m},\rho } \quad & \theta [ U^f(c_{1,f},l_{1,f})]+(1-\theta)[ U^m(c_{1,m},l_{1,m})] \\
& +\beta \Big[ (1-\pi) (\theta \times U^f(c_{2,f},0)+(1-\theta)U^m(c_{2,m},0)) \\
& +\pi E[\theta \times U^f(\alpha \times s,0)+(1-\theta)U^m((1-\alpha)s,0)] \Big] \\
\text{s.t.} \quad & c_{1,f} + c_{1,m} + s \leq (w_{1,f} l_{1,f} + w_{1,m} l_{1,m}) + S_0 \\
& c_{2,f} + c_{2,m} \leq s \quad \text{(If married in period 2)} \\
&\alpha=\rho  \quad \text{(Under TBR people share in case of divorce }\\
& \text{is equal to share of property under spouses name.)}
\end{align*}

\textbf{Notation:}
\begin{itemize}
    \item $c_{t,j}$ - consumption of spouse $j$ in period $t$
    \item $w_{t,j}$ - wage of spouse $j$ in period $t$
    \item $l_{t,j}$ - labor supply of spouse $j$ in period $t$
    \item $s$ - total household savings.
    \item $S_0$ - initial wealth
    \item $\pi$ is the probability of divorce
    \item $\theta$ is the bargaining power of wife
    \item $\beta$ is time discount factor.
    \item $\alpha$  - wife's share of savings in case of divorce
\end{itemize}
It can be translated to the following Lagrangian

\begin{multline}\label{eq: Lagrangian function}
    L=W+\lambda_1 \Big(w_{1,f} l_{1,f} + w_{1,m} l_{1,m}) + S_0-c_{1,f} - c_{1,m} - s \Big )+\lambda_2(s-c_{1,f}-c_{1,m})
\end{multline}

FOC:

\begin{equation}
    \frac{\partial L}{\partial c_{1,f}}=\theta U^f_c(c_{1,f},l_f)-\lambda_1=0
\end{equation}

\begin{equation}
    \frac{\partial L}{\partial c_{1,m}}=(1-\theta) U^m_c(c_{1,m},l_m)-\lambda_1=0
\end{equation}

\begin{equation}
    \frac{\partial L}{\partial c_{2,f}}=\beta (1-\pi) \theta U^f_c(c_{2,f},0)-\lambda_2=0
\end{equation}

\begin{equation}
    \frac{\partial L}{\partial c_{2,m}}=(1-\theta)\beta (1-\pi) U^m_c(c_{2,m},0)-\lambda_2=0
\end{equation}

\begin{equation}
    \frac{\partial L}{\partial \alpha}=\pi\theta U^f_c(\alpha s,0)s-\pi (1-\theta)U^m_c((1-\alpha) s,0)s =0
\end{equation}

\begin{equation}
    \frac{\partial L}{\partial s}=\pi\theta U^f_c(\alpha s,0) \alpha+\pi (1-\theta)U^m_c((1-\alpha) s,0)(1-\alpha) -\lambda_1+\lambda_2=0
\end{equation}

\begin{equation}
     \frac{\partial L}{\partial l_f}=\theta U^f_l(c_{1,f},l_f)+\lambda_1 w_f=0 
\end{equation}
\begin{equation}
     \frac{\partial L}{\partial l_m}=(1-\theta) U^m_l(c_{1,m},l_m)+\lambda_1 w_m=0 
\end{equation}

By examining the FOCs for consumption for both spouses in each period, I derive the following relationships:
\begin{equation}\label{eq: FOC c1f and c1m equal}
    \frac{U^f_c(c_{1,f},l_f)}{U^m_c(c_{1,m},l_m)}=\frac{1-\theta}{\theta}
\end{equation}

\begin{equation}\label{eq: FOC 2 period consumption combined}
    \frac{U^f_c(c_{2,f},0)}{U^m_c(c_{2,m},0)}=\frac{1-\theta}{\theta}
\end{equation}

\begin{equation}\label{eq: FOC 2 period consumption combined divorce}
    \frac{U^f_c(c_{2,f,d},0)}{U^m_c(c_{2,m,d},0)}=\frac{U^f_c(\alpha s,0)}{U^m_c((1-\alpha) s,0)}=\frac{1-\theta}{\theta}
\end{equation}

Given that $c_{2,f}+c_{2,m}=s$, WLOG I can say that 
\begin{equation}
    c_{2,f}=q s
\end{equation}

\begin{equation}
    c_{2,m}=(1-q) s
\end{equation}

So, instead of consumption $c_{2,f}$ and $c_{2,m}$, the household may choose a share in case of staying married, denoted as $q$, and savings, denoted as $s$. 

We can rewrite Equation \ref{eq: FOC 2 period consumption combined} as 

\begin{equation}\label{eq: FOC 2 period married in term of q}
     \Tilde{u}(q,s)= \frac{U^f_c(q s,0)}{U^m_c([1-q]s,0)}=\frac{1-\theta}{\theta}
\end{equation}

To produce the relationship between $q$ and $\alpha$, we combine Equations \ref{eq: FOC 2 period married in term of q} and \ref{eq: FOC 2 period consumption combined divorce}:

\begin{equation}\label{eq: relationship q alpha}
    \Tilde{u}(q,s)=\Tilde{u}(\alpha,s)
\end{equation}

\textbf{Proposition 1:}
\textit{Assume that:}
\begin{itemize}
    \item $U^j_c(0,l)=\infty$
    \item $U_c^j(c,l)>0$ - \textit{monotonous preferences}
    \item $U_{cc}^j(c,l)<0$ - \textit{decreasing marginal utility.}
\end{itemize} 

\textit{The household fully insures against divorce shock, i.e.,} $c_{2,j}=c_{2,j,d} \quad \forall j \in \{f,m\}$.

\subsubsection*{Proof:}
A sufficient condition for this equality to hold is the monotonicity of $\Tilde{u}(x,s)$. It is easy to show that the following holds:

\begin{equation}\label{eq: partial u}
    \frac{\partial \Tilde{u}(x,s)}{\partial x}=\frac{s\times U^f_{cc}(x s,0)U^m_c([1-q]s,0) +s U^m_{cc}([1-q]s,0) U^f_{c}(x s,0)}{\Big(U^m_c([1-q]s,0)\Big)^2}<0
\end{equation}

This holds if 

\begin{itemize}
    \item $U^j_c(0,l)=\infty$, which leads to $s>0$,
    \item $U_c^j(c,l)>0$ - monotonous preferences,
    \item $U_{cc}^j(c,l)<0$ - decreasing marginal utility.
\end{itemize}

Because $s>0$ and $U_c^j(c,l)>0 \quad \forall c, l$.

\textbf{Proposition 2: (Sufficient condition for EDR to increase (decrease) savings incentives)} 
\textit{EDR leads to higher savings if:}
\begin{itemize}
    \item a) 
    \begin{equation}\label{eq: threshold for Proposition 2}
    (1-\alpha^*) \times RP^f(\alpha^* s)+\alpha^*\times RP^m((1-\alpha^*) s) > (<) 2 
    \end{equation}
    \item b) 
    \begin{equation}
    \frac{U^f_{cc}(\alpha^* s,0)}{U^m_{cc}((1-\alpha^*) s,0)} = \frac{1-\theta}{\theta}\times \frac{1-\alpha^*}{\alpha^*}
    \end{equation}
\end{itemize}

\textit{where $\alpha^*$ is the optimal share of the wife in the case of a divorce, which the household would choose under TBR.}

\textit{and} 
\begin{equation}
    RP^j(x) = -x\frac{U_{ccc}^j(x,0)}{U_{cc}^j(x,0)} \quad \text{- relative prudence}
\end{equation}

\textbf{Proof: }
For sufficiency, consider cases when any deviation of the wife's divorce share, $\alpha$, from the optimal level under TBR would lead to an increase (decrease) in $\frac{\partial L}{\partial s}$. From FOC I know that 
\begin{equation}
   \frac{\partial L}{\partial s}(\alpha^*)=0
\end{equation}

To ensure that $\alpha=\alpha^*$ is a minimum (maximum), the following conditions must be met:
\begin{enumerate}
    \item \begin{equation}\label{eq: condition a}
    \frac{\partial^2 L}{\partial s\partial \alpha}(\alpha^*)=0
    \end{equation}
    \item \begin{equation}\label{eq: condition b}
    \frac{\partial^3 L}{\partial s\partial \alpha^2}(\alpha^*) > (<)0
    \end{equation}
\end{enumerate}

Considering \textbf{condition a)},
\begin{multline}
    \frac{\partial^2 L}{\partial s\partial \alpha}(\alpha^*)=\pi\theta U^f_{cc}(\alpha^* s,0) s \alpha^* +\pi\theta U^f_{c}(\alpha^* s,0) \\
    -s\pi (1-\theta)U^m_{cc}((1-\alpha^*) s,0)(1-\alpha^*) -\pi (1-\theta)U^m_{c}((1-\alpha^*) s,0)=0
\end{multline}

Using the FOC, which stipulates that
\begin{equation}
     \theta U^f_{c}(\alpha^* s,0)= (1-\theta)U^m_{c}((1-\alpha^*) s,0)
\end{equation}

we can simplify $\frac{\partial^2 L}{\partial s\partial \alpha}(\alpha^*)$ as
\begin{equation}
\boxed{
    \frac{U^f_{cc}(\alpha^* s,0)}{U^m_{cc}((1-\alpha^*) s,0)}=\frac{1-\theta}{\theta}\times \frac{1-\alpha^*}{\alpha^*}
    }
\end{equation}

Now, moving to \textbf{condition b)}, let us evaluate the following expression:
\begin{multline}\label{eq:  dL/(dsdada)}
    \frac{\partial^3 L}{\partial s \partial \alpha^2}(\alpha^*) =
    \pi s \Bigg [ \theta U^f_{cc}(\alpha^* s,0) \left( \frac{U^f_{ccc}(\alpha^* s,0)}{U^f_{cc}(\alpha^* s,0)} \alpha^* s +2 \right) +\\
    (1-\theta) U^m_{cc}((1-\alpha^*) s,0) \left( (1-\alpha^*)s \frac{U^m_{ccc}((1-\alpha) s,0)}{U^m_{cc}((1-\alpha^*) s,0)}+2 \right)\Bigg ]
\end{multline}

The relative prudence is an essential measure used in the literature to study precautionary savings motives, defined as:
\begin{equation}
    RP(x)=-x\frac{u'''(x)}{u''(x)}
\end{equation}

Using \(RP\), we can rewrite Equation \ref{eq:  dL/(dsdada)} as
\begin{multline}\label{eq:  dL/(dsdada) with RP}
    \frac{\partial^3 L}{\partial s \partial \alpha^2} =\pi s \Bigg \{ \theta U^f_{cc}(\alpha s,0) \left [2-RP^f(\alpha s) \right] +\\
    (1-\theta) U^m_{cc}((1-\alpha) s,0) \left [2-RP^m((1-\alpha^*)s) \right] \Bigg\} >( <)0
\end{multline}

Utilizing condition (a), and dividing Equation \ref{eq:  dL/(dsdada) with RP} by \(U^m_{cc}((1-\alpha) s,0)\) (keeping in mind that \(U^m_{cc}((1-\alpha) s,0) < 0\) due to concavity, thus necessitating an inversion of the inequality), we get:
\begin{multline}
    \frac{\partial^3 L}{\partial s \partial \alpha^2}(\alpha^*) \frac{1}{U^m_{cc}((1-\alpha^*)s,0)}=\\
    =\pi s (1-\theta) \frac{1}{\alpha^*}\Bigg \{ (1-\alpha^*)\left [2-RP^f(\alpha^* s) \right] +\alpha^* \left [2-RP^m((1-\alpha^*)s) \right] \Bigg\} <(>)0
\end{multline}

This expression is proportional to 
\begin{equation}
    2-(1-\alpha^*)RP^f(\alpha^* s) -\alpha^*RP^m((1-\alpha^*) s) <(>)0
\end{equation}
or, alternatively,
\begin{equation}
\boxed{
    (1-\alpha^*)RP^f(\alpha^* s) +\alpha^*RP^m((1-\alpha^*) s) >( <)2
    }
\end{equation}

\textit{Note:} Strictly speaking, the precautionary savings mechanism does not guarantee an increase in labor because the household might enhance savings by reducing total consumption in the initial period. Hence, in the present version, we only consider cases where the family finances increased savings through an augmentation of the labor supply.

\subsection{Special case: CRRA and $\gamma_f=\gamma_m=\gamma$}\label{sec: CRRA condition}

Given that the Constant Relative Risk Aversion (CRRA) utility function is utilized within the model, it's crucial to comprehend how propositions derived for the general case can be mirrored in our model. 

\subsubsection*{EDR and Precautionary Savings}

In this section, I demonstrate that in the case of a CRRA utility function, the sufficient condition also becomes a necessary one, which means that \textbf{Proposition 2} is valid in both directions. 

\textbf{Proposition 2a (CRRA case with \(\gamma_f = \gamma_m = \gamma\))}.

\textit{If} 
\[ U^j(c,l) = \frac{c^{1-\gamma}}{1-\gamma} - \phi_j l, \]
\textit{then Equilibrium Domestic Return (EDR) leads to an increase (decrease) in savings \textbf{if and only if}}
\begin{equation}
    \gamma >(<) 1.
\end{equation}

\textbf{Proof:} 
To prove \textbf{Proposition 2a}, it suffices to demonstrate that condition \textit{a} (Equation \ref{eq: condition a}) always holds, which inherently implies condition \textit{b} (Equation \ref{eq: condition b}).
\begin{equation}
    \frac{\partial^2 L}{\partial s\partial \alpha}(\alpha^*) = 0
    \label{eq: condition a restated}
\end{equation}

Let's define the utility function and its second derivative with respect to \(c\) as
\begin{equation}
    U^j(c,l) = \frac{c^{1-\gamma}}{1-\gamma} - \phi_j l
    \label{eq: utility function}
\end{equation}
\begin{equation}
    U^j_{cc}(c,l) = -\frac{\gamma(1-\gamma)}{c^{1+\gamma}}
    \label{eq: second derivative utility}
\end{equation}

The partial derivative of \(L\) with respect to \(s\) and \(\alpha\) at \(\alpha^*\) can be rewritten as
\begin{multline}
    \frac{\partial^2 L}{\partial s \partial \alpha}(\alpha^*) = \pi\theta \Big(-\frac{\gamma(1-\gamma)}{ (\alpha s)^{1+\gamma}}\Big) s \alpha^* - \\ -s\pi (1-\theta)\Big (-\frac{\gamma(1-\gamma)}{([1-\alpha^*]s)^{1+\gamma}}\Big)(1-\alpha^*) \\
    =\frac{\pi \gamma (1-\gamma)}{s^\gamma} \Bigg [  \frac{(1-\theta)}{[1-\alpha^*]^{\gamma}} - \frac{\theta}{(\alpha^*)^{\gamma}}\Bigg]
    \label{eq: partial derivative L}
\end{multline}

From Equation \ref{eq: FOC 2 period consumption combined divorce} , we determine \(\alpha^*\) as
\begin{equation}
    \frac{U^f_c(\alpha s,0)}{U^m_c((1-\alpha) s,0)}= \frac{(1-\alpha^*)^{\gamma}}{(\alpha^*)^{\gamma}} =\frac{1-\theta}{\theta}
    \label{eq: alpha determination}
\end{equation}

\begin{equation}
    \alpha^*=\frac{\theta^{\frac{1}{\gamma}}}{\theta^{\frac{1}{\gamma}}+(1-\theta)^{\frac{1}{\gamma}}}
    \label{eq: alpha star}
\end{equation}

It can be straightforwardly illustrated that
\begin{equation}
    \frac{\partial^2 L}{\partial s\partial \alpha}\left(\frac{\theta^{\frac{1}{\gamma}}}{\theta^{\frac{1}{\gamma}}+(1-\theta)^{\frac{1}{\gamma}}}\right) = 0
    \label{eq: demonstrated partial}
\end{equation}

With relative prudence being constant for CRRA:
\begin{equation}
    RP(c) = 1 + \gamma
    \label{eq: RP}
\end{equation}

The Equation \ref{eq: threshold for Proposition 2} simplifies to:
\begin{equation}
    \gamma >(<) 1
    \label{eq: simplified threshold}
\end{equation}
\hfill \(\square\)

\clearpage
\section{Robustness Checks}\label{sec A: robusness checks}
This section presents two robustness checks to ensure that the main results, from the main body of the paper, hold under more realistic assumptions. Note that these robustness checks are simulations of different models based on estimation results produced for the main model in Table \ref{table: estimation results}. 

\subsection{Endogenous Divorce}

The main model necessitates two modifications:

\begin{itemize}
    \item \textbf{``Love" Shocks.} In the second period, each spouse, $j$, experiences "love" shocks, $\xi_j$, during the marriage. 
    \item \textbf{Choice of Marital Status.} After observing the realization of "love" shocks, spouses decide whether they want to get divorced or stay married. Under Mutual Consent, a family can get a divorce only if both spouses agree to it.
\end{itemize}

These modifications alter the welfare function, $W$, as follows:

\begin{align*}
    \text{Period 1: } W_1 = & \theta \left[ u(c_{1,f}) -\phi_f l_{1,f}\right] +(1-\theta) \left[ u(c_{1,m})  -\phi_m l_{1,m}\right] + \beta E W_2\\
    \text{Period 2: } W_2 = & \left[ \theta \left(u(c_{2,f})+\xi_f\right)+(1-\theta)\left(u(c_{2,m})+\xi_m\right) \right] (1-d)\\
    & + d  \left[ \theta u(c_{2,f,d}) +(1-\theta)u(c_{2,m,d}) \right]
\end{align*}

where 
\[
    (\xi_f, \xi_m) \sim 
    N\left( 
    \begin{bmatrix} \mu_f^\xi \\ \mu_m^\xi \end{bmatrix}, 
    \begin{bmatrix} \sigma_f^\xi & 0 \\ 0 & \sigma_m^\xi \end{bmatrix} 
    \right)
\]
For simplicity, I assume that shocks are not correlated. $d=1$ if both spouses decide to get divorced: $u(c_{2,f})+\xi_f<u(c_{2,f,d})$ \textbf{AND} $u(c_{2,m})+\xi_m<u(c_{2,m,d})$. During the first period, the household expects to get divorced with probability, $\Tilde{\pi}$, where 
\begin{multline}
    \Tilde{\pi}= P\Big(u(c_{2,f})+\xi_f<E[u(c_{2,f,d})],u(c_{2,m})+\xi_m<E[u(c_{2,m,d})]\Big)=\\P\Big(u(c_{2,f})+\xi_f<E[u(c_{2,f,d})]\Big)\times P\Big(u(c_{2,m})+\xi_m<E[u(c_{2,m,d})]\Big)
\end{multline}

The parameters of the "love" shocks distribution are manually chosen so that the endogenous probability of divorce, $\Tilde{\pi}$, matches the exogenous one, $\pi$, from the main model. I utilize the expected value for utility from divorce because under EDR the share of property (savings) allocated to the wife is random.

\begin{figure}[h!]
    \centering
    \caption{Treatment effect size VS  risk-aversion,$\gamma$. Robustness check 1}
    \label{fig:ATT gamma: EQ-ttl-endog} 
    
    \begin{subfigure}{0.3\textwidth}
        \caption{ ) Wife's Weekly  Hours}
        \includegraphics[width=\linewidth]{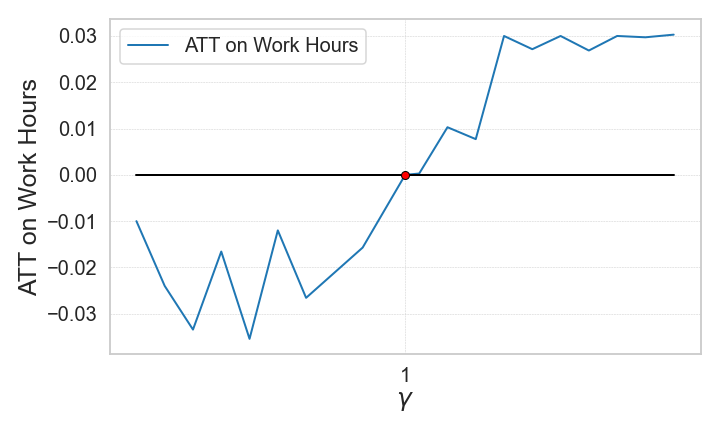}
        
    \end{subfigure}\hfill
    \begin{subfigure}{0.3\textwidth}
        \caption{ ) Husband's Weekly  Hours}
        \includegraphics[width=\linewidth]{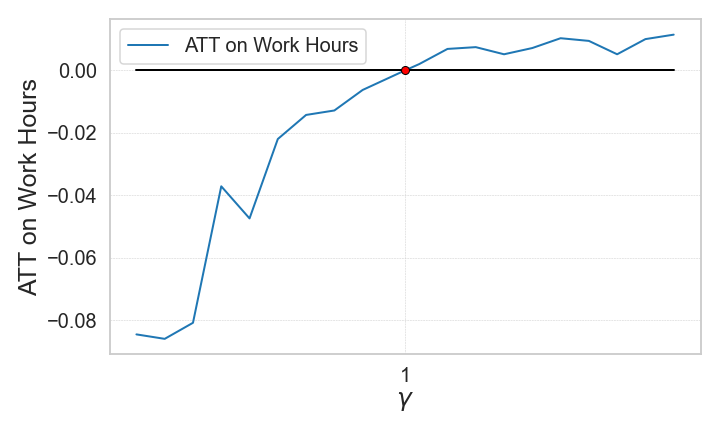}
        
    \end{subfigure}\hfill
    \begin{subfigure}{0.3\textwidth}
        \caption{) Weekly change in savings }
        \includegraphics[width=\linewidth]{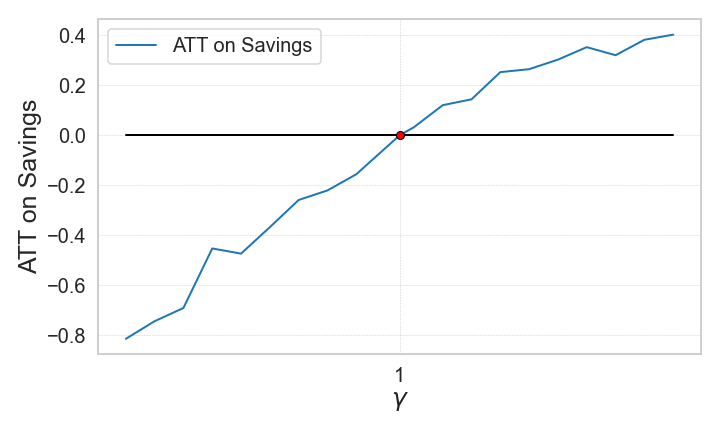}
        
    \end{subfigure}
     \caption*{\footnotesize \textit{Note}:  a) The  plots illustrate how the treatment effect (ATT) on labor supply and savings varies with $\gamma$. b) ATT is used to denote the difference in labor supply, $E\Big[l_j(EDR,\gamma) - l_j(TBR,\gamma)\Big]$, and in savings,$E\Big[s(EDR,\gamma) - s(TBR,\gamma)\Big]$, under different regimes.}
\end{figure}

Subsequently, I perform the same exercise as in section \ref{subsec: ATT-gamma}, with results displayed in Figure \ref{fig:ATT gamma: EQ-ttl-endog}.

As seen in Figure \ref{fig:ATT gamma: EQ-ttl-endog}, introducing endogenous divorce does not alter the results from the main body of the paper.

\subsection{Endogenous Divorce and "Bribing"}

The model is further extended by assuming that spouses can get divorced even if only one spouse desires it, by "bribing" the other spouse. For instance, if the wife desires a divorce as $u(c_{2,f})+\xi_f<E[u(c_{2,f,d})]$ where $c_{2,f,d}=\alpha s$, she might try to bribe the husband by choosing such a share of husband, $1-\Tilde{\alpha}$, which would make the husband indifferent: $u(c_{2,m})+\xi_m=u(\Tilde{c}_{2,m,d})=u((1-\Tilde{\alpha})s)$, while still keeping her inclined to get divorced: $u(c_{2,f})+\xi_f \leq u(\Tilde{c}_{2,f,d})$. Note that if the spouses can agree on a share, the court does not step in, so there is no uncertainty about the share, and expectation is dropped.

\begin{itemize}
    \item Case 1: $ \Bigg\{u(c_{2,f})+\xi_f\geq E[u(c_{2,f,d})] \quad \text{and} \quad u(c_{2,m})+\xi_m\geq E[u(c_{2,f,m})]\Bigg\}$ - couple stay married

    \item Case 2: $\Bigg\{u(c_{2,j})+\xi_j<E[u(c_{2,j,d})] \quad \text{and} \quad u(c_{2,-j})+\xi_{-j}>E[u(c_{2,-j,d})]    \Bigg\}$ 
    \begin{itemize}
        \item Case 2.1: $\Big \{ \exists \Tilde{\alpha} :\quad u(c_{2,-j})+\xi_{-j}=u(\Tilde{c}_{2,-j,d}) \quad \text{and} \quad   u(c_{2,j})+\xi_{j}<u(\Tilde{c}_{2,j,d})] \Big \}$ - couple get divorced, $\alpha=\Tilde{\alpha}$
        \item Case 2.2: $\Big \{ \nexists \Tilde{\alpha} :\quad u(c_{2,-j})+\xi_{-j}=u(\Tilde{c}_{2,-j,d}) \quad \text{and} \quad   u(c_{2,j})+\xi_{j}<u(\Tilde{c}_{2,j,d}) \Big \}$ - stay married.
    \end{itemize}
    \item Case 3: $\Bigg\{u(c_{2,j})+\xi_j<E[u(c_{2,j,d})] \quad \text{and} \quad u(c_{2,-j})+\xi_{-j}<E[u(c_{2,-j,d})]    \Bigg\}$ get married, $\alpha$ is random
\end{itemize}

\begin{figure}[h!]
    \centering
    \caption{Treatment effect size VS  risk-aversion,$\gamma$. Robustness check 2}
    \label{fig:ATT gamma: EQ-ttl-endog-bribe} 
    
    \begin{subfigure}{0.3\textwidth}
        \caption{ ) Wife's Weekly  Hours}
        \includegraphics[width=\linewidth]{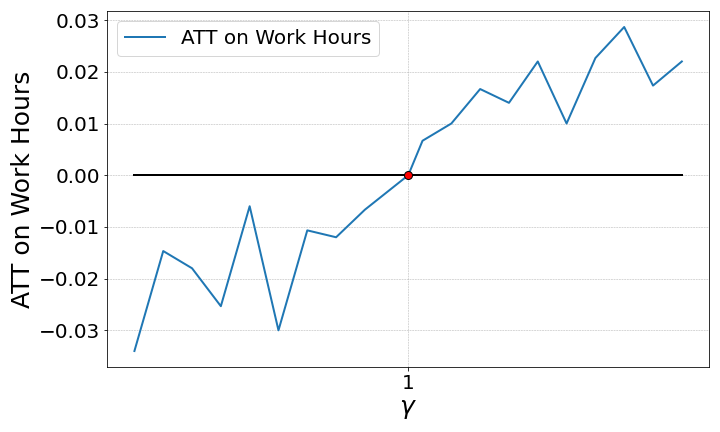}
        
    \end{subfigure}\hfill
    \begin{subfigure}{0.3\textwidth}
        \caption{ ) Husband's Weekly  Hours}
        \includegraphics[width=\linewidth]{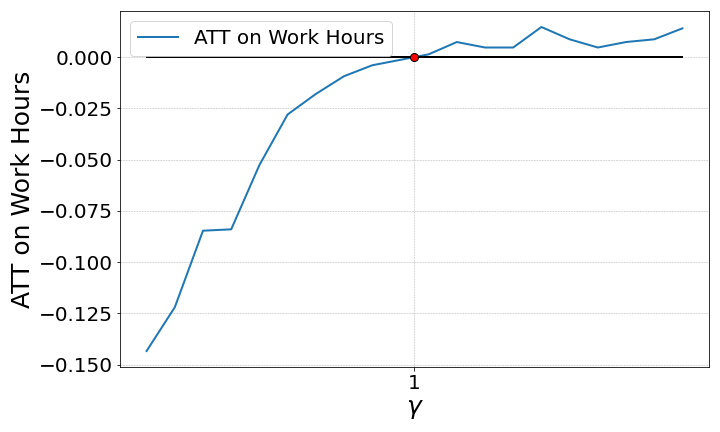}
        
    \end{subfigure}\hfill
    \begin{subfigure}{0.3\textwidth}
        \caption{) Weekly change in savings }
        \includegraphics[width=\linewidth]{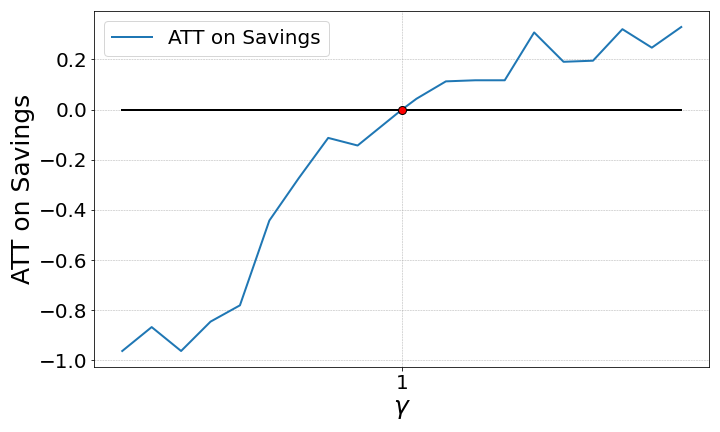}
        
    \end{subfigure}
     \caption*{\footnotesize \textit{Note}:  a) The  plots illustrate how the treatment effect (ATT) on labor supply and savings varies with $\gamma$. b) ATT is used to denote the difference in labor supply, $E\Big[l_j(EDR,\gamma) - l_j(TBR,\gamma)\Big]$, and in savings,$E\Big[s(EDR,\gamma) - s(TBR,\gamma)\Big]$, under different regimes.}
\end{figure}
\subsection{Non-binding Labor Supply.}

One might be concerned by the fact that on Figure \ref{fig:ATT gamma: EQ-ttl} both spouses work maximum number of hours and it does allow to see how policy would affect labor supply for people with $\gamma<1$. Because the hypotheses is that $\gamma$ is the only parameter which can affect the sign of treatment effect on labor supply. However, that comparative statics is not informative it labor supply constrain is binding. 
\clearpage
\section{Adoption of Law.}
\begin{table}[h!]
\centering
\caption{Divorce law reforms in the sample period}\label{table:divorce_law_reforms}
\begin{tabular}{lcc|lcc}
\toprule
State & U. Divorce & E. Distribution & State & U. Divorce & E. Distribution \\
\midrule
AL & 1971 & 1984 & MT & 1973 & 1976 \\
AK & pre-1967 & pre-1967 & NE & 1972 & 1972 \\
AZ & 1973 & CP & NV & 1967 & CP \\
AR & no & 1977 & NH & 1971 & 1974 \\
CA & 1970 & CP & NJ & no & CP \\
CO & 1972 & 1972 & NM & pre-1967 & 1980 \\
CT & 1973 & 1973 & NY & no & 1981 \\
DE & 1968 & pre-1967 & NC & no & 1981 \\
DC & no & 1977 & ND & 1971 & pre-1967 \\
FL & 1971 & 1980 & OH & 1992 & 1981 \\
GA & 1973 & 1984 & OK & pre-1967 & 1975 \\
HI & 1972 & pre-1967 & OR & 1971 & 1971 \\
ID & 1971 & CP & PA & no & 1980 \\
IL & no & 1977 & RI & 1975 & 1981 \\
IN & 1973 & pre-1967 & SC & no & 1985 \\
IA & 1970 & pre-1967 & SD & 1985 & pre-1967 \\
KS & 1969 & pre-1967 & TN & no & pre-1967 \\
KY & 1972 & 1976 & TX & 1970 & CP \\
LA & no & CP & UT & 1987 & pre-1967 \\
ME & 1973 & 1972 & VT & no & pre-1967 \\
MD & no & 1978 & VA & no & 1982 \\
MA & 1975 & 1974 & WA & 1973 & CP \\
MI & 1972 & pre-1967 & WV & 1984 & 1985 \\
MN & 1974 & pre-1967 & WI & 1978 & CP \\
MS & no & 1989 & WY & 1977 & pre-1967 \\
MO & no & 1977 & & & \\
\bottomrule
\end{tabular}

\footnotesize{Notes: Table adapted from \textcite{voena2015yours}. "CP" denotes Community Property regime.\\ It's similar to EDR; all property is pooled and split equally between spouses \\ regardless of ownership title.}
\end{table}

\end{appendix}

\end{document}